\documentclass{aa}  

\usepackage{tablefootnote}
\usepackage{graphicx, color}
\usepackage[space]{grffile}
\usepackage{latexsym}
\usepackage{float}
\usepackage{booktabs}

\usepackage{amsfonts,amsmath,amssymb}
\usepackage{url}
\usepackage{fancyref}
\usepackage{hyperref}
\usepackage{multirow}
\usepackage{txfonts}

\begin{document} 

\title{The Cosmic Radio Dipole: Estimators and Frequency Dependence} 

\subtitle{}
   \titlerunning{Quadratic Estimator}

\author{Thilo M.~Siewert\inst{1}\thanks{t.siewert@physik.uni-bielefeld.de} \and Matthias~Schmidt-Rubart\inst{1}\thanks{matthiasr@physik.uni-bielefeld.de}\and Dominik J.~Schwarz\inst{1}\thanks{dschwarz@physik.uni-bielefeld.de}}
	\authorrunning{Siewert et al.}

\institute{$^1$Fakult\"at f\"ur Physik, Universit\"at Bielefeld, Postfach 100131, 33501 Bielefeld, Germany \\
         }

\date{\today}

\abstract{The Cosmic Radio Dipole is of fundamental interest to cosmology. Recent studies revealed open questions about the nature of the observed Cosmic Radio Dipole.
We use simulated source count maps to test a linear and a quadratic estimator for its possible biases in the estimated dipole amplitude with respect to the masking procedure. 
We find a superiority of the quadratic estimator, which is then used to analyse the TGSS-ADR1, WENSS, SUMSS, and NVSS radio source catalogues, spreading over a decade of frequencies. 
The same masking strategy is applied to all four surveys to produce comparable results.
In order to address the differences in the observed dipole amplitudes, we cross-match the two surveys located at both ends of the analysed frequency range.
For the linear estimator, we identify a general bias in the estimated dipole directions.
The positional offsets of the quadratic estimator to the CMB dipole for skies with $10^7$ simulated sources is found to be below one degree and the absolute accuracy of the estimated dipole amplitudes is better than $10^{-3}$. 
For the four radio source catalogues, we find an increasing dipole amplitude with decreasing frequency, which is consistent with results from the literature and results of the cross-matched catalogue.
We conclude that for all analysed surveys, the observed Cosmic Radio Dipole amplitudes exceed the expectation, derived from the CMB dipole, which can not be explained by a kinematic dipole alone.}

\maketitle

\section{Introduction}\label{sec:introduction}

Starting from the Cosmological Principle, assuming an isotropic 
and homogeneous Universe, a prominent observed dipole anisotropy 
in the Cosmic Microwave Background (CMB) is assumed to be 
of kinematic nature. Possible other contributions from the 
integrated Sachs-Wolfe effect and primordial fluctuations 
are expected to be sub-dominant. As the assumption of an
isotropic Universe should also hold true for its matter 
distribution, observing visible matter should show a similar 
effect of proper motion.

For radio galaxies, the proper motion of the Solar 
system affects the observed source counts, which was first 
predicted by \citet{EllisBaldwin1984} and is often 
described as Cosmic Radio Dipole.
Several attempts have been made to measure the Cosmic Radio 
Dipole in terms of radio continuum surveys, such as the 
NRAO VLA Sky Survey (NVSS; \citealt{NVSS1998}) and 
Westerbork Northern Sky Survey (WENSS; \citealt{WENSS2000}).
In order to measure the vector of Cosmic Radio Dipole a 
variety of estimators have been introduced, like linear 
estimators using only the position of the sources
\citep{Crawford2009, Singal2011, GibelyouHuterer2012, RubartSchwarz2013} or estimators using source counts in 
spherical harmonics \citep{BlakeWall2002, TiwariNusser2016}.
The results of these different estimators and surveys spread 
from with CMB consistent measurements to several times the 
amount of amplitude. On the other hand, the measured 
dipole directions are not in tension with the CMB dipole.

Contributions from local structure and Poisson noise are 
expected to change the measured Cosmic Radio Dipole, whereas 
the structure induced dipole is expected to be dominant for 
sources limited to a redshift of $z\ll1$. 
\citet{NusserTiwari2015} and \citet{TiwariNusser2016} presented 
a model of the dipole component of NVSS radio galaxies, 
for which about $~70\%$ of the observed dipole signal, 
after removing the solar motion, is due to structures within 
$z \lesssim 0.1$. However, besides contributions from
cosmic structure and Poisson noise, the dipole due to our 
proper motion should be dominant for a catalogue with a mean 
redshift of order unity and large sky coverage.  
The aim of this work is to revisit the properties of linear and quadratic dipole estimators and to present a unified analysis for four wide area continuum radio surveys across a decade in frequency, in order to reveal the nature of the cosmic radio dipole.

After introducing the kinematic dipole in 
Sect.\ \ref{sec:cosmicradiodipole}, we discuss possible 
biases of the estimated dipole direction and amplitude 
for a linear estimator in Sect.\ \ref{sec:linearestimator}.
We compare them to the properties of a quadratic 
estimator, that was introduced in \citet{Bengaly2019} in 
Sect.\ \ref{sec:quadraticestimator}. The quadratic 
estimator is applied to four different radio surveys, 
which are described and masked in Sect.\ \ref{sec:Data}, 
and show the results of our dipole estimates in 
Sect.\ \ref{sec:results}. In Sect.\ \ref{sec:Discussion} 
we test and discuss the frequency dependence of the 
dipole amplitude. We conclude in Sect.\ \ref{sec:conclusions}.
Earlier studies and the estimators used in these studies 
are described in App. \ref{sec:otherresults}.

In this work 
we denote the equatorial coordinate system as right ascension 
$RA$ and declination $DEC$. When we use spherical coordinates, 
the azimuthal angle is written as $\vartheta = 90\mathrm{~deg} -DEC$
and the right ascension as $\varphi$.

\section{Cosmic Radio Dipole}\label{sec:cosmicradiodipole}

From the results of CMB analysis, the hypothesis 
$d_{cmb}=d_{motion}$ for the dipole amplitude is made. 
To test this, the dipole is measured by radio point 
source catalogues, where one expects
\begin{equation}
d_{\text{radio}}=d_{\text{motion}}+d_{\text{structure}}+d_{\text{noise}}
\end{equation}
Following \citet{EllisBaldwin1984} one can formulate the effect of 
a peculiar motion to the observed source counts of radio 
galaxies.

The emitted, or measured flux $S$ of an observed radio source follows a power law with spectral index $\alpha$,
\begin{equation}
S\propto \nu^{-\alpha}.
\end{equation}
While the spectral index can not be measured directly by 
most existing radio surveys, we assume a mean spectral 
index for all sources. In general, the spectral index can vary 
between sources, depending on the sourc type and the most 
prominent radiation process in the source.
In a study comparing radio sources of the NVSS and the 
TIFR GMRT Sky Surveys first alternative data release (TGSS-ADR1; 
\citealt{TGSS2017}) an averaged $\bar{\alpha} = 0.7870 \pm 0.0003$
\citep{Gasperin2018} was found. 
Using the same radio surveys, with additional criteria on the 
radio sources, \cite{Tiwari2016} estimated a mean spectral 
index of $\bar{\alpha}= 0.763\pm 0.211$.
For simplicity and consistency to earlier studies 
\citep{RubartSchwarz2013}, we use a spectral index of $\alpha=0.75$.

The observed number of radio sources per solid angle is expected 
to follow a simple power law as function of flux density,
\begin{equation}\label{eq:sourcedistribution}
\frac{\text{d}N}{\text{d}\Omega}(>S)\propto S^{-x},
\end{equation}
with slope parameter $x$. Most commonly the slope is assumed to 
be $x=1$. 

Assuming a proper motion of velocity $\mathrm{v}$ between the 
observer and the cosmic rest frame, two effects have to 
be taken into account. The first effect is that the observed frequency $\nu_{obs}$ is Doppler,
\begin{equation}
\nu_{\text{obs}}=\nu_{\text{rest}}(1+\beta\cos\theta)\gamma,
\end{equation}
with $\theta$ the angle between the line of sight and the 
direction of motion, $\beta=\mathrm{v}/c$ and 
$\gamma=1/\sqrt{1-\beta^2}$.
The finite speed of light leads to aberration and thus changes 
the position of each source towards the direction of motion. 
The angle $\theta$ between the direction of motion and the 
position of the source changes like
\begin{equation}\label{eq:aberration}
\tan\theta_{obs}=\frac{\sin\theta_{rest}}{\beta+\cos\theta_{rest}}\gamma^{-1}.
\end{equation}
With this new angle, the equal area of Eq. (\ref{eq:sourcedistribution}) changes in first order of $\beta$ to
\begin{equation}
d\Omega_{obs}=d\Omega_{rest}\left(1-2\beta\cos\theta\right)+\mathcal{O}(\beta^2).
\end{equation}
Combining both effects leads to the predicted number density
\begin{equation}\label{eq:boostedcounts}
\left(\frac{\text{d}N}{\text{d}\Omega}\right)_{obs}=\left(\frac{\text{d}N}{\text{d}\Omega}\right)_{rest}\left[1+[2+x(1+\alpha)]\beta\cos\theta\right]+\mathcal{O}(\beta^2).
\end{equation}
The change in source counts or density is then given by the amplitude $d$ of the kinematic radio dipole and can be written as 
\begin{equation}\label{eq:dipoleamplitude}
d=[2+x(1+\alpha)]\beta.
\end{equation}

Assuming a spectral index $\alpha=0.75$, slope $x=1$, and 
the latest findings of the Planck CMB study \citep{PlanckA2018},
which found a kinematic dipole velocity of 
$\mathrm{v}=369.82 \pm 0.11$~km/s towards 
$(167.942\pm0.011,\ -6.944\pm0.005)$~deg (Equatorial coordinates, J2000), we find an expected kinematic radio dipole amplitude of
\begin{equation}
d_{CMB}= 0.46\times10^{-2}.
\end{equation}

\section{Linear estimator}\label{sec:linearestimator}

In \citet{RubartSchwarz2013} the dipole amplitude measured by linear estimators like 
\begin{equation}
\label{eq:DefLinearEst}
 \vec R_{\mathrm{3D}} = \sum \vec{\hat r}_i 
\end{equation}
was discussed in great detail, with special attention toward the correction of an amplitude bias. This estimator was proposed by \citet{Crawford2009} and later used e.g. in \citet{Singal2011} or \citet{Singal2019}.

The dipole direction on the other hand was assumed to be unbiased, as long as the masked map was point symmetric w.r.t. the observer. This assumption was first made in the work of \cite{EllisBaldwin1984} and explicitly used in later studies, e.g. by \cite{Singal2011}. 
The basic principle is that the monopole will not appear in the linear estimation, as long as the mask is symmetric, since the monopole contribution due to masking will cancel then.
Implicitly the same assumption is used, when masked areas are filled with isotropically distributed simulated sources, for correcting the monopole bias for incomplete skies \citep{Crawford2009,GibelyouHuterer2012}. 
In fact this assumption is over-simplifying the problem, since the radio sky does also have a dipole modulation (plus higher order multipoles), which can interact with the mask. In this general case, linear estimators like (\ref{eq:DefLinearEst}) will obtain a directional bias too.

One expects the biggest possible bias to emerge, when the mask is not symmetric with respect to the dipole modulation. 
We created simulated radio source maps and used the linear estimator to measure the dipole direction. 
In this way we can compare, whether the measured direction is in agreement with the true simulated distribution.
We simulated the sky with one million isotropically distributed sources and modified this with a kinematic dipole modulation, using a velocity of $v=1200$~km/s towards $RA = 180$~deg and $DEC_d^{\rm{sim}}$, the number count of the simulated sources behaved like $n(>S) \propto S^{-x}$ with $x=1$ and the spectral index of the sources was $\alpha=0.75$, leading to an expected dipole amplitude of $d=1.5 \times 10^{-2}$. (see Table \ref{tab:MaskBias}).
We masked areas in two different ways. The first type of mask is called "caps", because the areas at the polar caps were masked and only sources between the declination values of $40$~deg and $-40$~deg were included in the measurement. 
Exactly inverted is the mask type "ring", where only sources outside those declinations will be taken into account. 

\begin{table}
\centering
\caption{Linear estimator measurements of dipole amplitude ($d$) and dipole declination ($DEC_d$) for 100 simulated maps containing $10^6$ sources (for the full sky) and an observer velocity of $v=1200$~km/s towards $RA = 180$~deg and $DEC_d^{\rm{sim}}$, the source count of the simulated sources behaves like $n(>S) \propto S^{-x}$ with $x=1$ and the spectral index of the sources is $\alpha=0.75$, leading to an expected dipole amplitude of $d=1.5 \times 10^{-2}$. Masking sources within $-40^\circ <\delta< +40$~deg for the ring type mask and with either $\delta>40$~deg or $\delta<-40$~deg for the caps type mask. }
\begin{tabular}{cccc}
\hline\hline
mask type & $DEC_d^{\rm{sim}}$ & measured $d$ & measured $DEC_d$ \\ 
& (deg) &($\times 10^{-2}$) & (deg)\\ \hline
caps	&   $-40$ & $1.565 \pm 0.017$ & $-14.95 \pm 0.44$ \\ 
ring	&   $-40$ & $2.016 \pm 0.026$ & $-74.60 \pm 0.38$ \\
caps	&   $-10$ & $1.917 \pm 0.024$  & $-3.21 \pm 0.38$ \\
ring	&   $-10$ & $0.915 \pm 0.018$ & $-35.84 \pm 1.30$ \\ \hline
\end{tabular}
\label{tab:MaskBias}
\end{table}

In Table \ref{tab:MaskBias} the results of the measured simulated dipole amplitude ($d$) and dipole declination ($DEC_d$) from our simulations can be seen. 
The changing dipole amplitude values were expected and for observations using radio surveys these amplitudes are therefore modified by a masking factor e.g. \citep{RubartSchwarz2013}. 
In all simulated cases we also see a directional bias significantly above the estimated variances of the simulations. 
For the caps mask, the bias goes towards the celestial equator and for the ring mask type towards the celestial poles (in this case, the celestial South Pole). 
Due to symmetry considerations the caps mask in general will have an effect pointing towards the equator and the ring mask away from the equator, independent of the sign of the dipole declination values.

For the simulated cases above we can calculate the direction bias analytically. 
Therefore we assume w.l.o.g. $N$ sources on the whole sky and a dipole with simulated right ascension of  $\varphi_d= 0$~deg.
The dipole amplitude $d$ and the dipole declination $\tilde{\vartheta}_d$ will not be fixed.
For convenience we will switch the coordinate system into spherical coordinates, meaning we change all declination angles into $\vartheta= 90$~deg$-\tilde{\vartheta}$.
We calculate the expectation value of the linear estimator with a mask that removes sources within the polar caps (up to an angular distance to the poles of $\psi$),
\begin{equation}
 \langle \vec{R}_{\mathrm{3D}} \rangle = \frac{N}{4 \pi}\int_{|\cos{\vartheta}|<|\cos{\psi}|} \mathrm{d} \Omega\,  (1 +  \hat{r} \cdot \vec{d})\, \hat{r} .
\end{equation}
We define $\vec{d}$ as a vector of amplitude $d$ pointing towards the dipole direction. With $\mu=\cos{\psi}$ the integral becomes:
\begin{equation}
 \frac{N}{4 \pi}\int_0^{2\pi} \mathrm{d}\varphi \int_{-\mu}^\mu \mathrm{d}\cos{\vartheta}\  \left[1 +  \left(\begin{array}{c} 
 \cos{\varphi} \sin{\vartheta} \\ 
 \sin{\varphi} \sin{\vartheta} \\  
 \cos{\vartheta}\end{array}\right)  \cdot d \left(\begin{array}{c} 
 \sin{\vartheta}_d \\ 
 0 \\  
 \cos{\vartheta}_d\end{array}\right) \right] \, \hat{r}  .
\end{equation}
Executing the scalar product, the integrand gives
\begin{equation}
 \curvearrowright \left[ 1+ d \cos{\varphi} \sin{\vartheta} \sin{\vartheta_d}+d \cos{\vartheta} \cos{\vartheta_d} \right]\ \left(\begin{array}{c} 
 \cos{\varphi} \sin{\vartheta} \\ 
 \sin{\varphi} \sin{\vartheta} \\  
 \cos{\vartheta}\end{array}\right) \ .
\end{equation}

The $y$ component of $\langle \vec{R}_{\mathrm{3D}} \rangle$ vanishes, since the corresponding terms in the integrand are proportional to $\sin{\varphi}$ or $\sin{\varphi}\cos{\varphi}$ and will therefore result in $0$ after the integration over $\mathrm{d}\varphi$. Now we evaluate the $z$ component. Here the terms $\cos{\vartheta} +d \cos{\vartheta}  \cos{\varphi} \sin{\vartheta} \sin{\vartheta_d}$ vanish after $\cos{\vartheta}$ and $\varphi$ integration respectively. Hence, we are left with
\begin{equation}
 \langle \vec{R}_{\mathrm{3D}} \rangle_z=\frac{N}{4 \pi}\int_0^{2\pi} \mathrm{d}\varphi \int_{-\mu}^\mu \mathrm{d}\cos{\vartheta} (d\cos^2{\vartheta}\cos{\vartheta_d}) \ .
\end{equation}
After resubstituting $\mu$, this integral provides the expectation value
\begin{equation}
 \langle \vec{R}_{\mathrm{3D}} \rangle_z=\frac{N}{3} d \cos{\vartheta_d} \cos^3{\psi} \ .
\end{equation}
Here we see that the masking limit $\cos{\psi}$ does have an effect on the  $z$ component of the estimator. In order to understand, whether the effect propagates to the direction estimation, we also have to evaluate the $x$ component of the integral. Here the only non vanishing term in the integrand leads to

\begin{equation}
 \langle \vec{R}_{\mathrm{3D}} \rangle_x=\frac{N}{4 \pi}\int_0^{2\pi} \mathrm{d}\varphi \int_{-\mu}^\mu \mathrm{d}\cos{\vartheta} (d\cos^2{\varphi}\sin^2{\vartheta}\sin{\vartheta_d}) \ ,
\end{equation}
which can be solved, utilising the relation $\sin^2{\vartheta}=1-\cos^2{\vartheta}$ and again changing $\mu$ back to $\cos{\psi}$, leading to:
\begin{equation}
 \langle \vec{R}_{\mathrm{3D}} \rangle_x=\frac{N}{2} d \sin{\vartheta_d} \left[ \cos{\psi-\frac{1}{3}\cos^3{\psi}} \right]  \ .
\end{equation}

The direction estimation $\vartheta_e$ can be obtained by $\tan{\vartheta_e}=\frac{\langle \vec{R}_{\mathrm{3D}} \rangle_x}{\langle \vec{R}_{\mathrm{3D}} \rangle_z}$, which leads to 

\begin{equation}
 \tan{\vartheta_e}=\frac{\langle \vec{R}_{\mathrm{3D}} \rangle_x}{\langle \vec{R}_{\mathrm{3D}} \rangle_z}=\frac{N/2\ d \sin{\vartheta_d} \left[ \cos{\psi-\frac{1}{3}\cos^3{\psi}} \right]}{N/3\ d \cos{\vartheta_d} \cos^3{\psi}} \ ,
\end{equation}
and this simplifies into
\begin{equation}\label{thetabiasinclude}
 \tan{\vartheta_e}= \tan{\vartheta_d} \frac{\left[3-\cos^2{\psi} \right]}{ 2 \cos^2{\psi}} \ .
\end{equation}
When applying no mask, meaning $\psi=0\mathrm{~deg} \rightarrow \cos{\psi}=1$, the direction estimation becomes unbiased, since the last factor in equation (\ref{thetabiasinclude}) will be unity.
For the case of a caps type mask, the behaviour will be described by the factor
\begin{equation}B_c= \frac{3-\cos^2{\psi}}  {2\cos^2{\psi}}.
\end{equation}
Before this result is compared to the simulation, we also consider the case of a mask excluding sources within a ring of $|\cos{\vartheta}|<|\cos{\psi}|$. For this we can utilize the previous calculation, since we are considering the exactly inverted case. Hence, the expectation of each component will be the value for the whole sky, with the values derived above subtracted, for sources within the ring, so
\begin{equation}
 \langle \vec{R}_{\mathrm{3D}} \rangle_z=\frac{N}{3} d \ \cos{\vartheta_d} \left[1- \cos^3{\psi} \right]
\end{equation}
and 
\begin{equation}
 \langle \vec{R}_{\mathrm{3D}} \rangle_x=\frac{N}{3} d \ \sin{\vartheta_d} \left[1-\frac{3}{2} \left(\cos{\psi}-\frac{1}{3}\cos^3{\psi} \right) \right] \ .
\end{equation}
From this we obtain
\begin{equation}\label{thetabiasexclude}
 \tan{\vartheta_e}= \frac{\langle \vec{R}_{\mathrm{3D}} \rangle_x}{\langle \vec{R}_{\mathrm{3D}} \rangle_z}=\tan{\vartheta_d} \frac{  1- \frac{3}{2}\cos{\psi+\frac{1}{2}\cos^3{\psi}} }{1-  \cos^3{\psi}} \ .
\end{equation}
This time the case of no applied mask corresponds to $\psi=90\mathrm{~deg} \rightarrow \cos{\psi}=0$ and the estimator becomes unbiased again. For both cases we obtain a different bias, but the behaviour is the same in principle. The tangent of the declination will be multiplied by a factor. 
 When we mask sources inside a ring, the bias factor becomes 
 \begin{equation}
 B_r=\frac{2- 3\cos{\psi}+\cos^3{\psi} }{2-  2\cos^3{\psi}}.
 \end{equation}

\begin{table}
\centering
\caption{Comparison of 
$|\tan{\vartheta_d}^{\rm{est}}|/|\tan{\vartheta_d}^{\rm{sim}}|$ and bias factors $B_{c/r}$ with simulated results from linear estimator measurements for 100 simulated maps with $10^6$ sources (for the full sky) and an observer velocity of $v=1200$~km/s towards $\varphi = 180$~deg and $\rm{tan}\vartheta_d^{\rm{sim}}$, the source count of the simulated sources behaves like $n(>S) \propto S^{-x}$ with $x=1$ and the spectral index of the sources is $\alpha=0.75$. Masking sources within $-40^\circ <\delta< +40$~deg for the ring type mask and with either $\delta>40$~deg or $\delta<-40$~deg for the caps type mask. }
\begin{tabular}{ccccc}
\hline\hline
mask type & $|\tan{\vartheta_d}^{\rm{sim}}|$ &  $|\tan{\vartheta_d}^{\rm{est}}|$ & $\frac{|\tan{\vartheta_d}^{\rm{est}}|}{|\tan{\vartheta_d}^{\rm{sim}}|}$ & $B_{c/r}$\\ \hline
caps	&   $1.19$ & $3.75 \pm 0.12$ & $3.14 \pm 0.11$ & $3.13$ \\ 
ring	&   $1.19$ & $0.28 \pm 0.01$ & $0.23 \pm 0.01$ & $0.23$ \\
caps	&   $5.67$ & $17.84 \pm 2.12$  & $3.15 \pm 0.38$& $3.13$ \\
ring	&   $5.67$ & $1.38 \pm 0.07$ & $0.24 \pm 0.01$ & $0.23$ \\ \hline
\end{tabular}
\label{tab:BiasMasktest}
\end{table}

In order to verify this derivation, we compared the calculated bias factors (see Table \ref{tab:BiasMasktest}) with the results from the simulations, shown in Table \ref{tab:MaskBias}. 
One can directly see that the bias factors from the simulations fit very well to the calculated ones within the estimated uncertainties. Hence, we understood this effect for both discussed cases and can conclude that masking a ring or masking areas outside a ring does have a significant effect on the dipole direction estimation in general.

The estimated dipole direction will be effectively pushed away from the masked areas. In both discussed cases the expectation value of the direction $\vartheta_e$ will be
\begin{equation}
 \vartheta_e= \arctan \left(B_{c/r}  \tan{\vartheta_d} \right) \ ,
\end{equation}
with $B_{c/r} $ standing for the bias factor $B_c$ or $B_r$, depending on the mask type. The total change in angle is small, when the distance between the masked areas and the dipole is big (meaning close to $90$~deg), since $\arctan$ becomes very flat for those angles. So a dipole direction, which is far away from the masked region is not biased as much and would therefore be comparably stable. 
For example, a masked ring with $\pm 10$ degrees and a Dipole direction at 45 degrees away from the ring (centre) leads to a change in estimated dipole direction of roughly one degree.

For the case of a mask consisting of one ring, (excluding or including sources within) the bias can be corrected for, since it is now fully understood. Unfortunately the mask, used in real dipole estimation, is more complicated (see Sect. \ref{sec:Masking} for more details). Each mask consists of at least two masking rings (with different orientations), which in general even overlap. Those cases cannot be handled as trivially by deriving a general bias factor. 

Summarized, we found that the dipole direction estimated from a very simple linear estimator is of limited use, since it has a directional bias, which cannot easily be corrected for in the general case. This limitation does not apply for more complicated linear estimators, like the ones utilizing spherical harmonics [e.g. \cite{Jain13}].

\section{Quadratic estimator}\label{sec:quadraticestimator}
In order to estimate the Cosmic Radio Dipole, we employ a quadratic estimator, which was shown first in \citet{Bengaly2019} and \citet{SKARedBook2018}. The estimator compares a model of source counts with dipole amplification per cell to the observed source counts per cell,
\begin{equation}\label{eq:estimator}
\chi^2=\sum_{i=0}^{N_{cell}}\frac{(N_{i,o}-N_{i,m})^2}{N_{i,m}}.
\end{equation}
The model is written as
\begin{equation}\label{eq:model}
N_{i,m}=m(1+\vec{d}\cdot \vec{e_i}),
\end{equation}
where $\vec{e_i}$ are unity vector towards the cell centres of the given data cells. The sum in the estimator is taken over all unmasked cells.
The estimation is done by minimizing $\chi^2$ in the parameter space of monopole and dipole amplitude, as well as possible dipole position. 
To get equal and non-overlapping cells, we use the pixelation scheme of {\sc HEALPix}\footnote{\url{http://healpix.sourceforge.net}} \citep{Healpix2005}, which was developed for the Cosmic Microwave Background.
The pixelation in {\sc HEALPix} is defined in the terms of a resolution parameter $N_{side}$, where the number of cells to cover the whole sky is defined as $N_{pix}=12N_{side}^2$.
The grid of possible dipole positions is generated by using {\sc HEALPix} cell positions in a same or higher resolution than the data.
The dipole amplitude is varied in the range $0.0\leq d \leq 0.1$ in steps of $\Delta d = 5\times 10^{-4}$. The monopole amplitude is estimated in a $5\%$ range around the mean number of sources of all cells with $6$ intermediate steps.

As for all other implemented estimators, it is necessary to mask unobserved regions, as well as regions with contributions from local structure, or systematical uncertainties.
Comparing the observed source counts map to a test sky, it is not necessary to have a symmetrical mask at all.
Therefore we perform a masking strategy consisting out of several steps. Not observed cells and cells within the Galaxy are masked and rejected in the dipole estimation. 
The mask itself is defined by a binary {\sc HEALPix} map, which is applied to the source count map.
Additionally criteria can easily be added to this mask.
So we reject pixels of sources rather than individual sources.

\begin{figure}
\centering
\includegraphics[width =\linewidth]{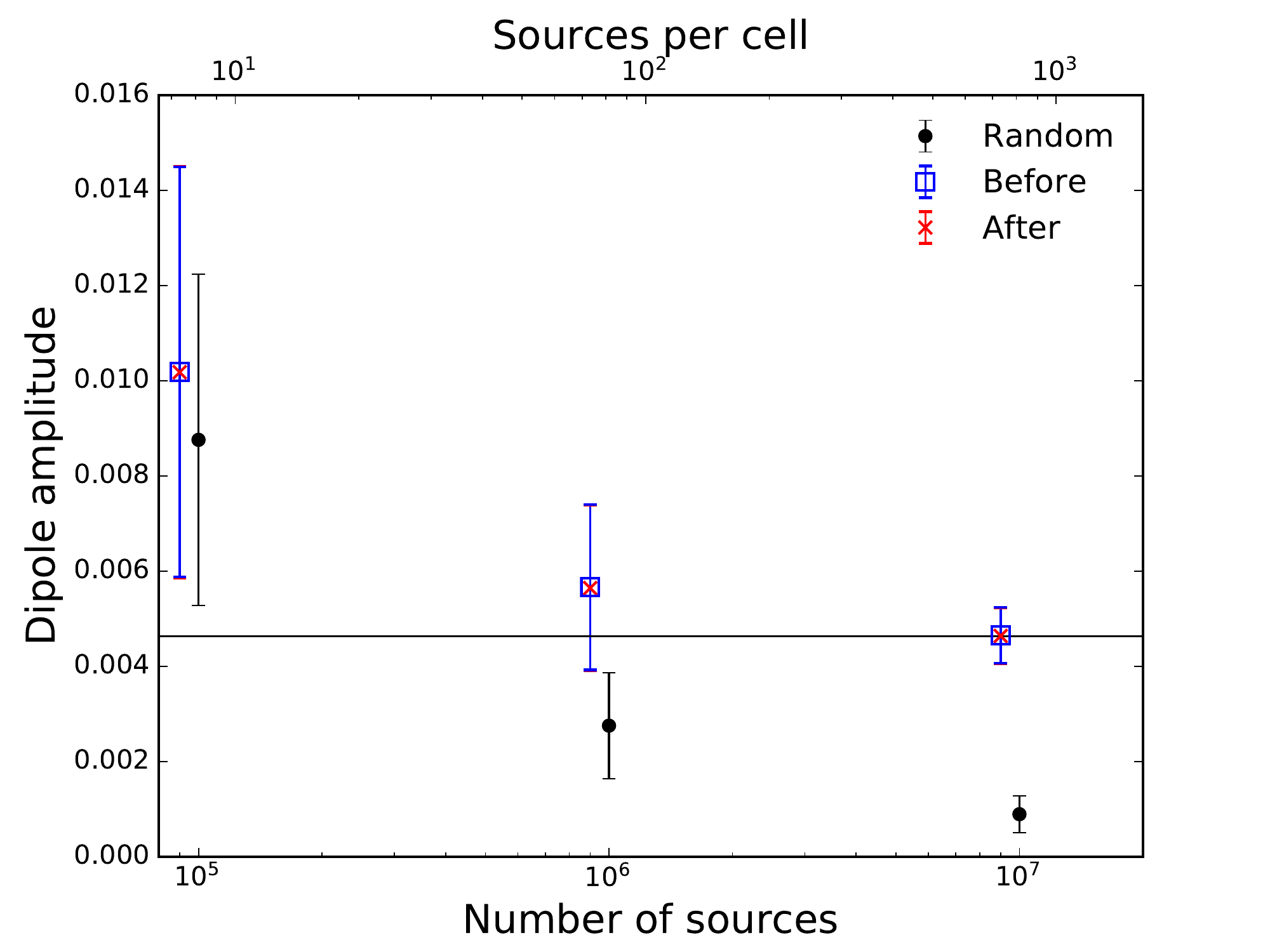}
\caption{Dipole amplitudes of the quadratic estimator for different sample sizes of purely random skies (black points), skies boosted before pixelization (blue boxes) and skies boosted after pixelization (red crosses). The solid line indicates the simulated amplitude $d= 4.63\times10^{-3}$. The simulations and the estimations are performed on a {\sc HEALPix} grid with resolution $N_{side}=16$. }
\label{fig:QETest_amp}
\end{figure}

\begin{figure}
\centering
\includegraphics[width =\linewidth]{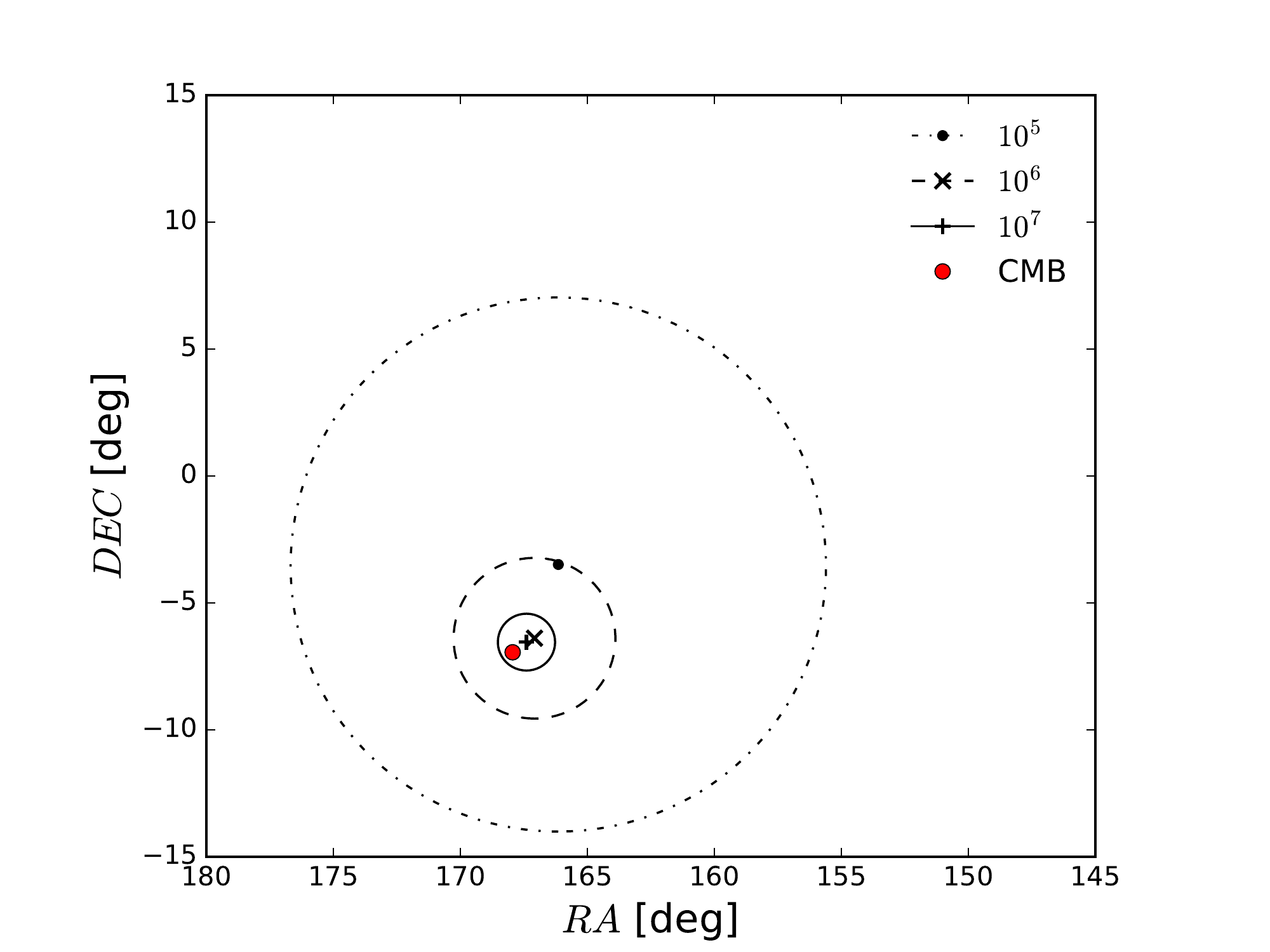}
\caption{Dipole directions (markers) of the quadratic estimator for different sample sizes of skies boosted after pixelization, shown in cartesian projection together with the $68\%$ confidence cones (lines). The red dot indicates the simulated dipole direction in equatorial coordinates $(RA,DEC)=(167.942,-6.944)$~deg. The simulations and the estimations are performed on a {\sc HEALPix} grid with resolution $N_{side}=16$. }
\label{fig:QETest_dir}
\end{figure}
\subsection{Simulations}
To test the quadratic estimator we simulate random skies with varying number of sources and try to recover the fiducial kinematic dipole.
In order to generate a random set of position on the sphere, we are using the default {\sc python} package \texttt{random}.
To uniformly draw random position on a sphere, we choose values in the range $[0,1)$ and transform them via:
\begin{align}
\varphi &= 360~\text{deg}\cdot \texttt{random(0,1)}\\
\vartheta &= \frac{180~\text{deg}}{\pi}\arccos(1-2\cdot\texttt{random(0,1)})
\end{align}
Using this definition, we already fulfil the convention of Co-Latitude necessary for {\sc HEALPix}.
Additionally we can generate random flux densities:
\begin{equation}
S = S_0\ (1-\texttt{random(0,1)})^{-1/x},
\end{equation}
which follow a source count distribution like presented in Eq. (\ref{eq:sourcedistribution}).
From this purely random set of positions we can infer dipole boosted mocks in two ways: Either by Doppler shifting and applying the effect of aberration to every source (see Sect. \ref{sec:cosmicradiodipole}) with a spectral index $\alpha = 0.75$ and number count slope $x=1$. Or by boosting the source counts per cell directly by applying Eq. (\ref{eq:model}) with an dipole amplitude $d_{CMB} = 0.46\times 10^{-2}$ and spectral index and number count slope from the former.
Here we do not make any assumption on clustering, different source populations or redshift dependencies.
The purely random samples are tested directly, whereas we apply a flux density threshold to the sources either before we boost the source counts per cell, or after we boost the sources directly.
This resembles sources close to an arbitrary detection limit of a real radio survey, which would be boosted over, or below the detection threshold.

The measurement is then computed as mean and error from 100 random samples, which is straightforward for the dipole amplitude.
Based on \citet{FisherLewis1983} and \citet{SphericalData} the mean direction of a spherical sample, with spherical coordinates $(\vartheta_i, \varphi_i)$ can be calculated via their euclidean coordinates $(x_i,y_i,z_i)$.
The euclidean coordinates for $\vartheta_i\in[0, \pi]$ and $\varphi_i\in [0, 2\pi]$ are:
\begin{align}
    x_i &= \sin\vartheta_i \cos\varphi_i\\
    y_i &= \sin\vartheta_i \sin\varphi_i\\
    z_i &= \cos\vartheta_i
\end{align}
The mean direction $(\bar\vartheta, \bar\varphi)$ of this sample can then be calculated via the mean cartesian coordinates:
\begin{align}
    \bar{S}_x &= \sum_i^n x_i/n,\text{ or } S_x = n\bar{S}_x\\
    \bar{S}_y &= \sum_i^n y_i/n,\text{ or } S_y = n\bar{S}_y\\
    \bar{S}_y &= \sum_i^n z_i/n,\text{ or } S_z = n\bar{S}_z\\
    \bar{R}^2 &= \bar{S}^2_x + \bar{S}^2_y + \bar{S}^2_z,\text{ or } R^2 = n^2\bar{R}^2
\end{align}
which leads to:
\begin{align}\label{eq:sphericalmean1}
    \bar\vartheta&= \arccos \left(\bar{z}/\bar{R}\right),\\\label{eq:sphericalmean2}
    \bar\varphi &= \arctan \left(\bar{y}/\bar{x}\right),
\end{align}
with $\bar\vartheta \in [0, \pi]$ and $\bar\varphi\in [0, 2\pi]$.
For large enough samples \citep[$n\geq25$]{FisherLewis1983,SphericalData} an approximated confidence cone, centred on $(\bar\vartheta, \bar\varphi)$ can be computed.
The confidence cone is characterised by the quantile $\alpha$, which corresponds to a $100\ (1-\alpha)\%$ confidence.
The semi-vertical angle $q$ of the sample confidence cone can then be calculated via:
\begin{equation}\label{eq:anglecone}
    q = \arcsin\left((-\log\alpha)^{1/2}\sigma\right)
\end{equation}
with the estimated spherical standard error $\sigma$:
\begin{align}
    \sigma &= \left(d/(n\bar{R}^2)\right)^{1/2},\\
    d &= 1-(1/n)\sum_i^n\left(x_iS_x+y_iS_y+z_iS_z\right)^2.
\end{align}

We use Eq. (\ref{eq:sphericalmean1}) and (\ref{eq:sphericalmean2}) to calculate the spherical mean direction of the 100 simulated skies, boosted with a fiducial kinematic dipole with velocity $\mathrm{v}= 370$~km/s ($d=0.46\times10^{-2}$), or $\mathrm{v}= 1200$~km/s ($d=1.51\times10^{-2}$).
The semi-vertical angle of the confidence cone, calculated using Eq. (\ref{eq:anglecone}), is compared to the offset $\Delta\theta_{CMB}$ between the spherical mean direction and the fiducial dipole direction $(\varphi_f,\vartheta_f)$ of the CMB (see Sect. \ref{sec:cosmicradiodipole}):
\begin{equation}\label{eq:offset}
\Delta\theta = \cos^{-1}{\left(\sin{(\vartheta_m)}\sin{(\vartheta_f)}+\cos{(\vartheta_m)}\cos{(\vartheta_f)}\cos{(\varphi_m-\varphi_f)}\right)}.
\end{equation}
The estimated dipole amplitudes from different sizes of simulated skies with fiducial CMB dipole and without are shown in Fig. \ref{fig:QETest_amp}, whereas the estimated dipole directions of simulated skies with fiducial CMB dipole are shown in Fig. \ref{fig:QETest_dir} with a $68\%$ confidence cone.
As the source density of existing radio surveys is about 10 to 100 sources per square degree, we restrict our tests to $10^5$, $10^6$ and $10^7$ sources for a full sky. 

Using possible survey properties from the upcoming Square Kilometre Array (SKA), the estimator of Eq. (\ref{eq:estimator}) was already briefly tested with simulations of the Cosmic Radio Dipole \citep{SKARedBook2018,Bengaly2019}.
Based on the simulated source counts with $10^8$ to $10^9$ sources, inferred from theoretical power spectra including clustering and evolution, the dipole position (in galactic coordinates) was estimated to have maximal errors of $(\Delta l, \Delta b) \sim (9,5)$~deg and with errors on the recovered dipole amplitude of $\sim 10\%$.
\begin{table}
    \centering
    \caption{Dependence of the results from the quadratic estimator on different searching grid resolutions for a simulated dipole amplitude of $d=0.46\times10^{-2}$. }
    \begin{tabular}{ccccccc}
         \hline\hline
         	\addlinespace[0.5ex]$N_{side}^{grid}$ & $N$ & $\overline{RA}$ & $\overline{DEC}$& $q_{68}$ & $q_{95}$ &  $\Delta\theta_{CMB}$\\
         && (deg)& (deg) & (deg) & (deg) & (deg)\\	\addlinespace[0.5ex]\hline
         \multirow{3}{*}{16} & $10^5$ & 166.14 & -3.49  & 10.52 & 17.22 & 3.89 \\
         & $10^6$& $167.13$ & $-6.42$ & 3.16 & 5.14 & 0.96 \\
        & $10^7$& 167.40 & -6.54  & 1.12 & 1.82 & 0.67 \\\hline
         \multirow{3}{*}{32} & $10^5$ & 165.98& -3.08  & 10.53 & 17.23 & 4.33  \\
         & $10^6$ & 167.25 & -6.54 & 3.16 & 5.12 & 0.79  \\
         & $10^7$ & 167.39 & -6.40  & 1.11 & 1.80 & 0.77\\\hline
         \multirow{3}{*}{64} & $10^5$ & 166.11 & -3.40  & 10.51 & 17.20 & 3.98 \\
         & $10^6$ & 167.20 & -6.49  & 3.15 & 5.12 &  0.87  \\
         & $10^7$ & 167.38 & -6.43 & 1.11 & 1.80 & 0.75\\\hline
    \end{tabular}
    \label{tab:gridsizeCMB}
\end{table}
\begin{table}
    \centering
    \caption{Dependence of the results from the quadratic estimator on different searching grid resolutions for a simulated dipole amplitude of $d=1.50\times10^{-2}$. }
    \begin{tabular}{ccccccc}
         \hline\hline
         	\addlinespace[0.5ex]$N_{side}^{grid}$ & $N$ & $\overline{RA}$ & $\overline{DEC}$& $q_{68}$ & $q_{95}$ &  $\Delta\theta_{CMB}$\\
         && (deg)& (deg) & (deg)& (deg)& (deg)\\	\addlinespace[0.5ex]\hline
         \multirow{3}{*}{16} & $10^5$ & 165.10 & -6.45  & 4.01 & 6.52 & 2.86 \\
         & $10^6$& 168.63 & -6.86 & 1.10 & 1.78 & 0.69 \\
        & $10^7$& 168.10 & -6.59  & 0.39 & 0.63 & 0.39 \\\hline
         \multirow{3}{*}{32} & $10^5$ & 165.30 & -6.31  & 3.99 & 6.48 & 2.70  \\
         & $10^6$ & 168.58 & -6.96 & 1.10 & 1.79 & 0.63  \\
         & $10^7$ & 168.02 & -6.64  & 0.35 & 0.57 & 0.32\\\hline
         \multirow{3}{*}{64} & $10^5$ & 165.33 & -6.24  & 3.98 & 6.47 & 2.69 \\
         & $10^6$& 168.61 & -6.95 & 1.09 & 1.77 & 0.67 \\
         & $10^7$ & 168.01 & -6.70 & 0.35 & 0.56 & 0.25\\\hline
    \end{tabular}
    \label{tab:gridsizestrong}
\end{table}
Over all sample sizes, the recovered position and amplitude of the fiducial kinematic dipole are comparable between boosting before and after the pixelization. 
Because of the high difference in computation time, we choose to boost the source counts directly in further studies.
In order to see a percent effect in the source count fluctuations, one has to overcome the shot noise based on the counting process in each cell. 
By measuring the dipole amplitude in a purely random sample, we can estimate a lower limit of sources necessary to see the fiducial simulated dipole, see Fig. \ref{fig:QETest_amp}. 
\subsection{Grid and masking effects}
To test if the estimated dipole position is affected by the resolution of the used grid, we test the same simulated skies with three different grid resolutions, namely $N_{side}=16,32$ and 64. 
The mean spacing of these resolutions between neighbouring cell centres is $\theta_{16}= 3.66$~deg, $\theta_{32}= 1.83$~deg and $\theta_{64}= 0.92$~deg, with number of possible directions (cell positions) $3072$, $12\,288$ and $49\,152$ respectively. 
In Table \ref{tab:gridsizeCMB} and \ref{tab:gridsizestrong} the estimated dipole directions with errors and offsets to the simulated dipole with $d=0.46\times10^{-2}$ and $d=1.51\times10^{-2}$ amplitudes are presented.
While using higher resolutions of the position grid we observe a decreased offset of the estimated dipole direction to the CMB dipole.
The error on the estimated dipole direction also decreases slightly with higher resolutions. 
As the time for one estimation increases together with the resolution, we decide for further analyses to use a position grid with $N_{side}^{grid}=32$. For the data we keep a pixelization resolution of $N_{side}=16$.

\begin{figure}
    \centering
    \includegraphics[width=0.9\linewidth]{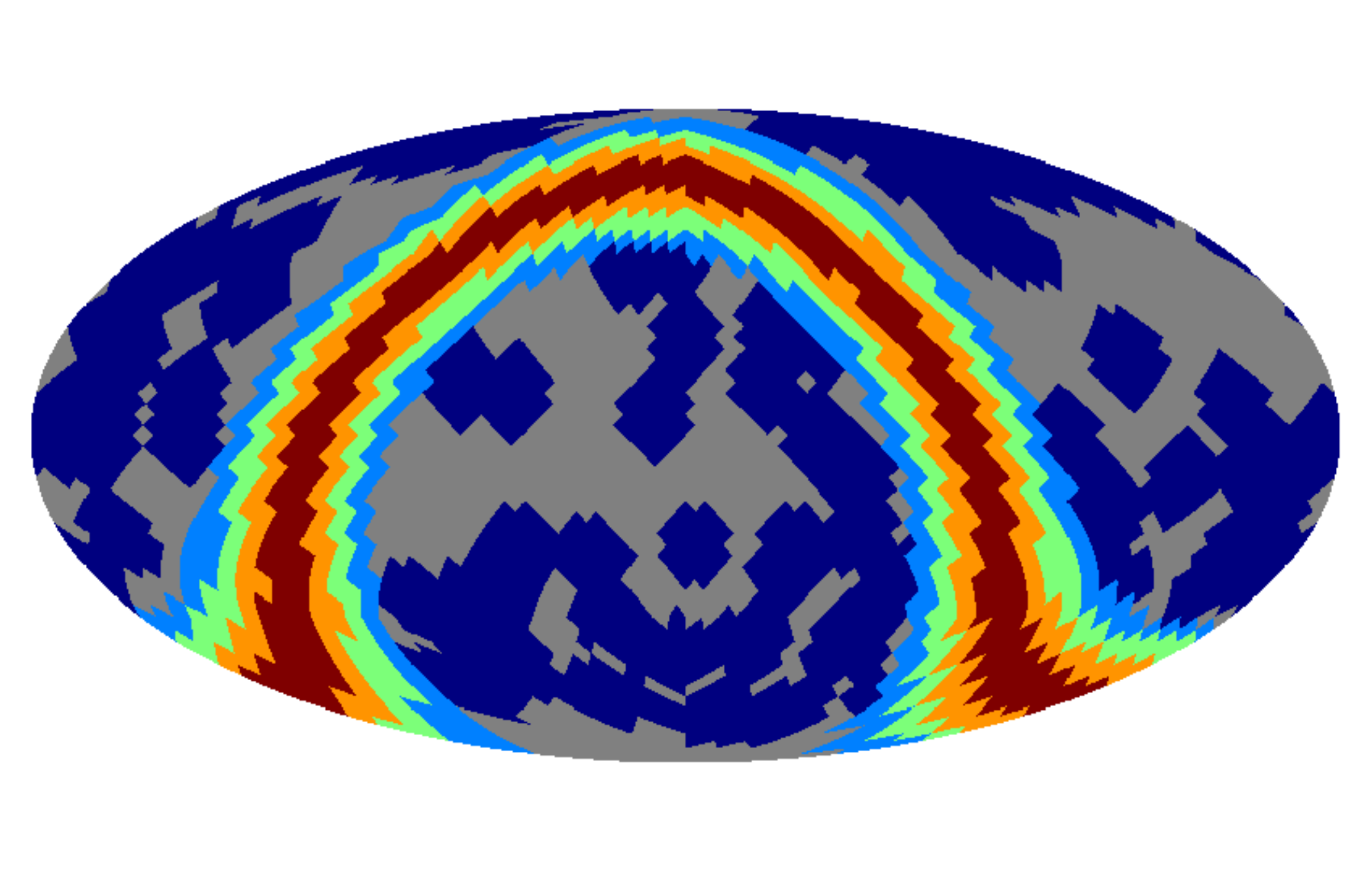}
    \caption{Example masks for different cuts in galactic latitude, with $|b|=5$~deg (red), 10~deg (orange), 15~deg (green) and 20~deg (light blue). Additionally 100 random discs with radius 10~deg are masked (blue). The unmasked region is highlighted in grey.}
    \label{fig:masks}
\end{figure}
\begin{figure}
    \centering
    \includegraphics[width=\linewidth]{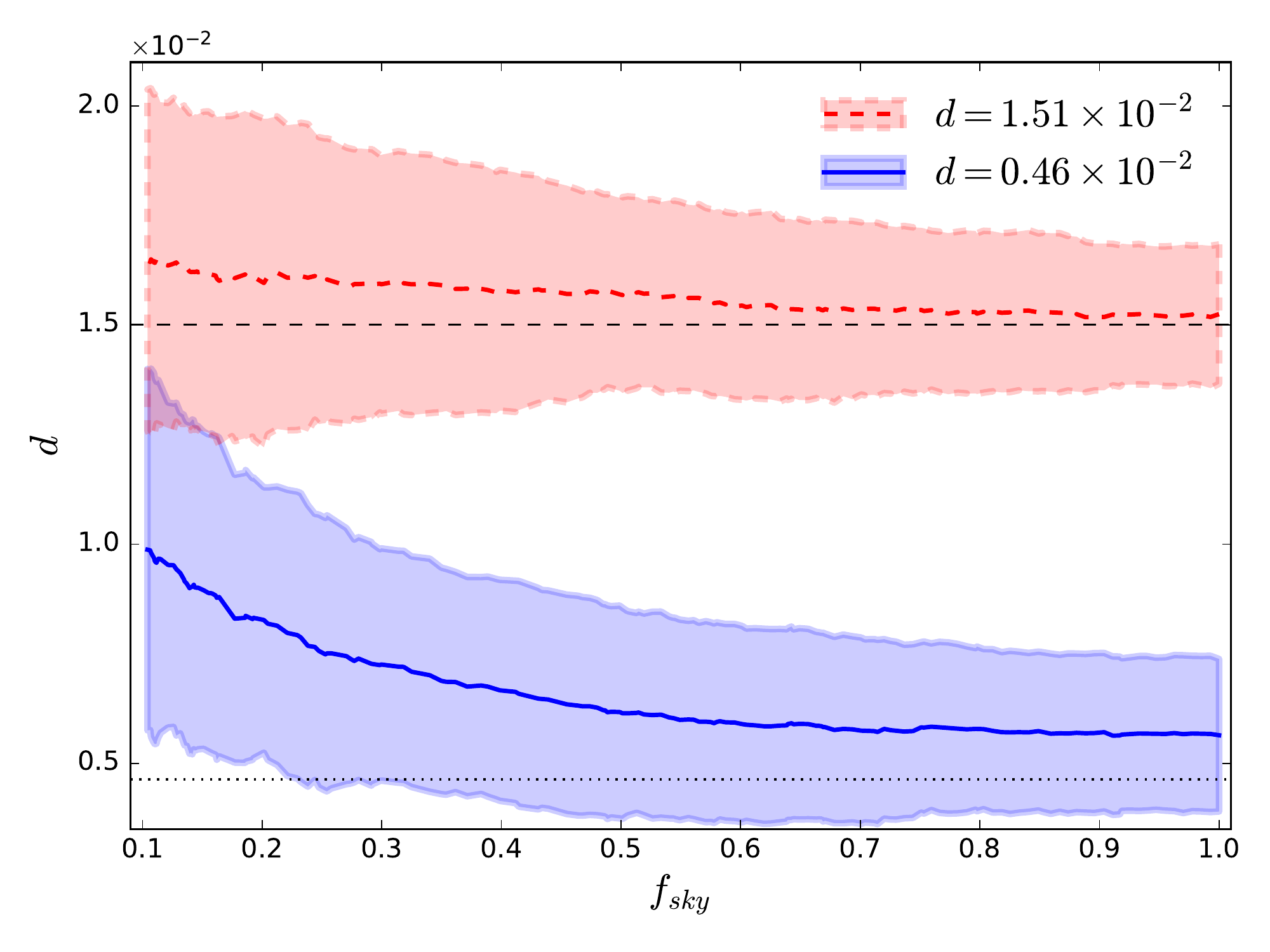}
    \caption{Estimated dipole amplitude for simulated skies with $d=1.51\times10^{-2}$ (dashed line) and $d=0.46\times10^{-2}$ (solid line). 
    The decreasing sky coverage was generated by masking 300 discs with 10~deg radius at random positions. The fiducial kinematic radio dipole amplitude is shown as dashed, or dotted line, respectively.}
    \label{fig:RandomCircles}
\end{figure}
\begin{figure}
    \centering
    \includegraphics[width =\linewidth]{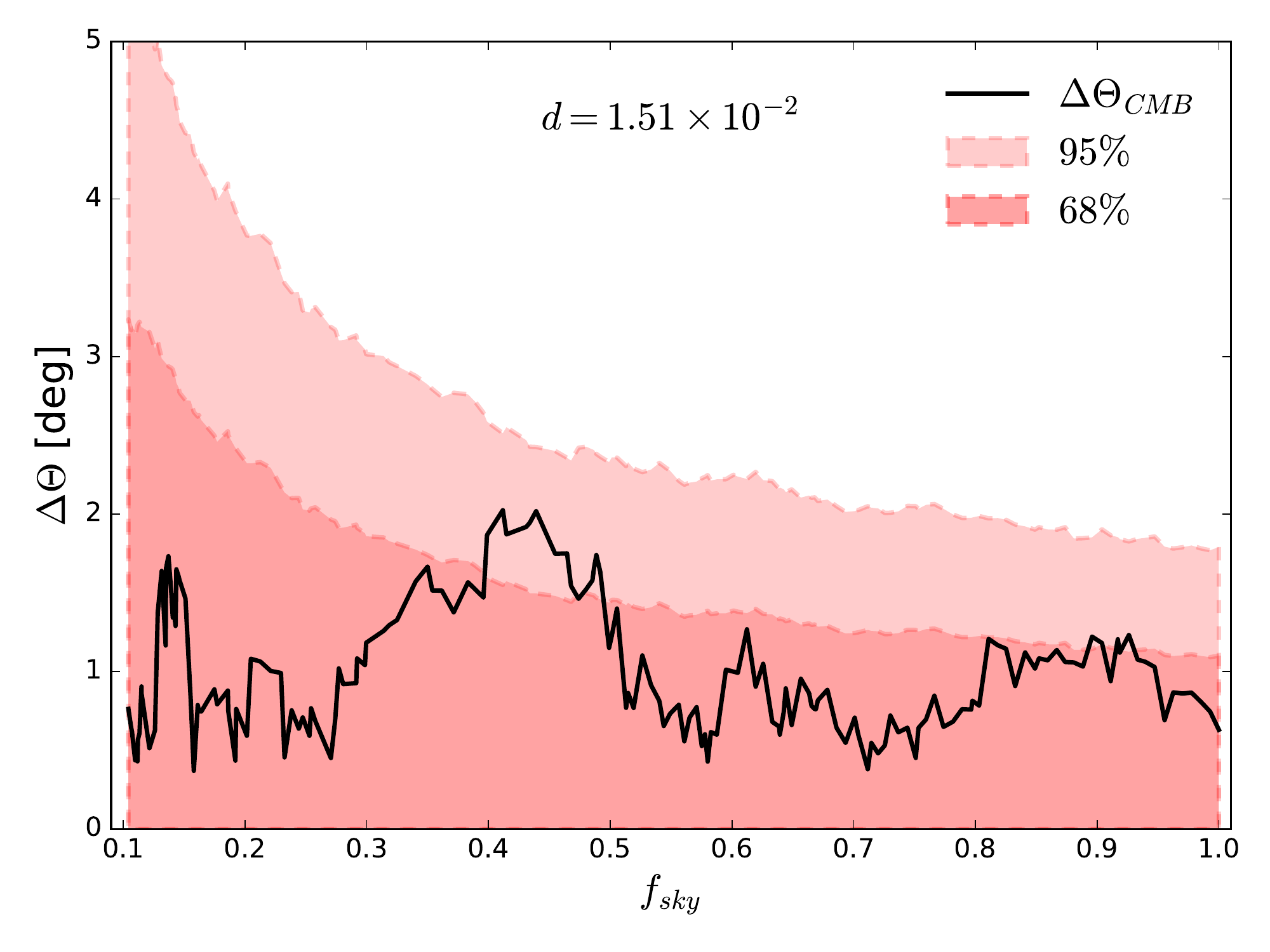}\\
    \includegraphics[width =\linewidth]{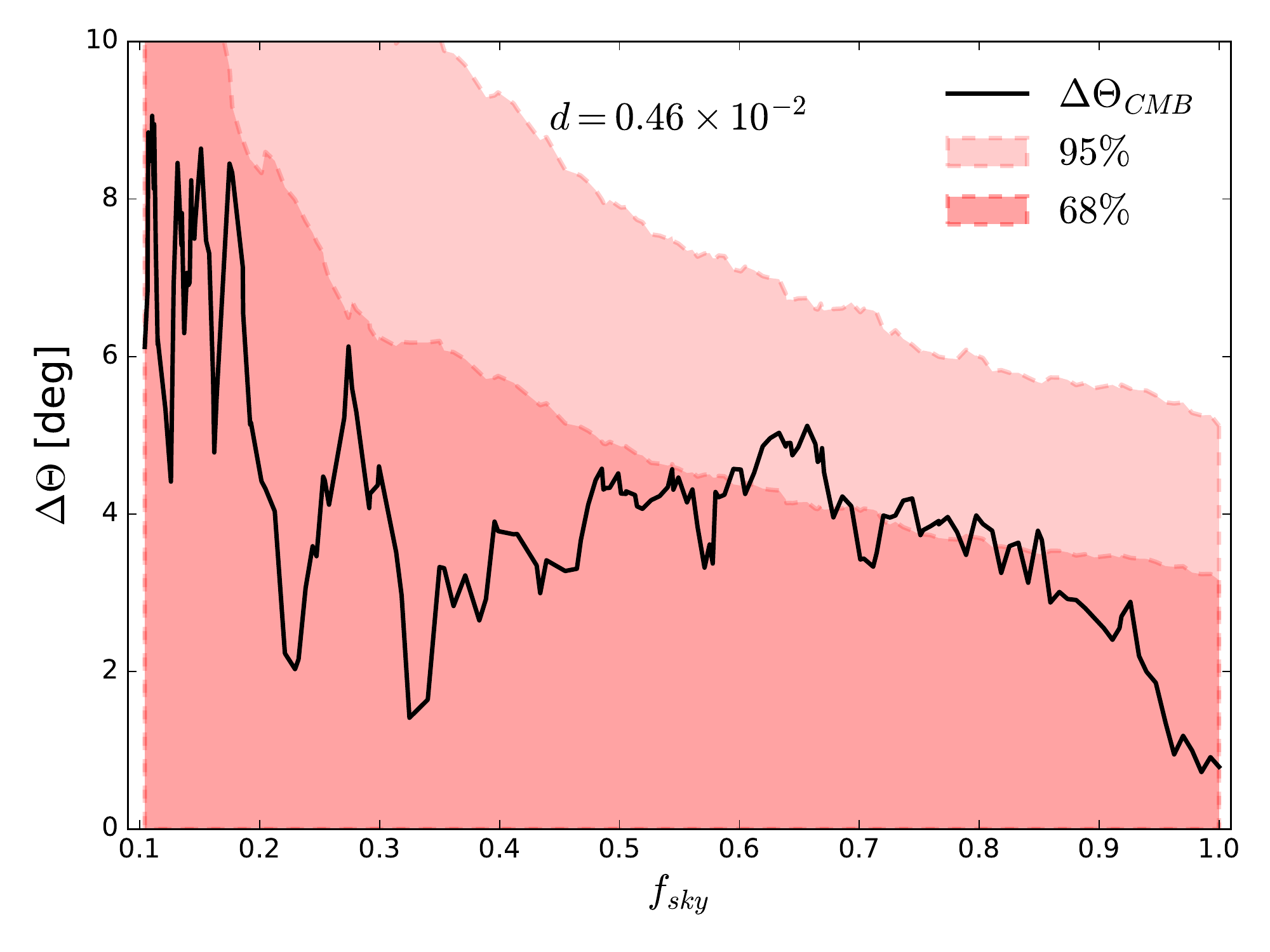}
    \caption{The semi-vertical angles of 
    the 68 and 95\% confidence cones (shaded) and the angular offset of the mean direction from the fiducial dipole direction (lines) for $d=1.51\times10^{-2}$ (top) and $d=0.46\times10^{-2}$ (bottom). The decreasing sky coverage was generated by masking 300 discs with 10~deg radius at random positions.}
    \label{fig:RCCone}
\end{figure}

Additionally we tested if the quadratic estimator is biased by possible masking procedures.
Therefore we use the boosted skies from the former and apply two different kind of masks to the simulations. 
Firstly, we test a galactic cut of equal latitude in steps of five degrees from 5 to 20~degrees.
The masks are shown in Fig. \ref{fig:masks} as red (5~deg), orange (10~deg), green (15~deg) and light blue (20~deg) regions, with the results of the estimations shown in Table \ref{tab:masks}. 
We observe an increasing positional offset, as well as an increasing positional error when masking larger galactic cuts. 
The estimated dipole directions with $10^6$ sources is for all masks within two degrees and for $10^7$ sources even below the spacing of the used position grid.
For all samples with $10^6$ and $10^7$ sources the estimated dipole amplitudes are in agreement with the simulated kinematic dipole, but increase slightly with increasing galactic cuts. 

\begin{table*}
    \centering
    \caption{Results of the quadratic estimator for masks with different galactic cuts ($|b|$) and total number of sources before masking, boosted with a kinematic dipole of $d=0.46\times10^{-2}$. We additionally show the results for the unmasked sky. }
    \begin{tabular}{cccccccc}
    \hline\hline
    \addlinespace[0.5ex]$|b|$ & $N$ & $\overline{RA}$ & $\overline{DEC}$& $q_{68}$ & $q_{95}$ &  $\Delta\theta_{CMB}$ & $d$\\
        (deg)&&(deg)& (deg) & (deg) & (deg) &(deg) &$(\times 10^{-2})$ \\\hline
        \multirow{3}{*}{-} &$10^5$ & 165.98& -3.08  & 10.53 & 17.23 & 4.33& $1.02\pm 0.43$\\
        & $10^6$& 167.25 & -6.54 & 3.16 & 5.12 & 0.79& $ 0.56\pm 0.17$ \\
        & $10^7$& 167.39 & -6.40  & 1.11 & 1.80 & 0.77 & $ 0.46\pm 0.06$ \\\hline
        \multirow{3}{*}{5} & $10^5$& 164.14 & -2.47 &10.69 & 17.50& 5.86 &  $1.08\pm 0.46$\\
        &$10^6$ &168.99& -6.14& 3.36&5.45 & 1.31 &  $0.57\pm 0.18$\\
        & $10^7$&167.40 & -6.48& 1.18 & 1.91 & 0.71 & $0.47\pm 0.06$\\\hline
        \multirow{3}{*}{10} & $10^5$ &165.24 & 0.55 & 11.83 & 19.42 & 7.96 & $1.12\pm 0.48$\\
        & $10^6$&167.62 & -4.52 & 3.65 & 5.92 & 2.44 & $0.59\pm 0.19$\\
        & $10^7$&167.31 & -5.91 & 1.28 & 2.08 & 1.21 & $0.47\pm 0.07$\\\hline 
        \multirow{3}{*}{15} & $10^5$ & 165.12 & 7.60 & 12.76 & 20.99 & 14.81 & $1.16\pm 0.44$ \\
        & $10^6$& 167.74 & -5.69 & 3.83 & 6.22 & 1.27  & $0.60\pm 0.19$\\
        & $10^7$ & 167.03 & -5.78 & 1.34 & 2.17 & 1.47 & $0.47\pm 0.08$\\\hline
        \multirow{3}{*}{20} & $10^5$ & 170.30 & 6.06 & 14.47 & 23.90 & 13.21  & $1.28\pm 0.47$\\ 
        & $10^6$& 167.87 & -5.34 & 4.13 & 6.71 & 1.60 & $0.62\pm 0.19$\\
        & $10^7$ &167.28 & -5.76 & 1.47& 2.39 & 1.36  & $0.47\pm 0.07$\\\hline
    \end{tabular}
    \label{tab:masks}
\end{table*}

\begin{table*}
    \centering
    \caption{Results of the quadratic estimator for masks with different galactic cuts ($|b|$) and total number of sources before masking, boosted with a kinematic dipole of $d=1.51\times10^{-2}$. We additionally show the results for the unmasked sky. }
    \begin{tabular}{cccccccc}
    \hline\hline
    \addlinespace[0.5ex]$|b|$ & $N$ & $\overline{RA}$ & $\overline{DEC}$& $q_{68}$ & $q_{95}$ &  $\Delta\theta_{CMB}$ & $d$\\
        (deg)&&(deg)& (deg) & (deg) & (deg) & (deg) &$(\times 10^{-2})$ \\\hline
        \multirow{3}{*}{-} & $10^5$ & 165.30& -6.31  & 3.99 & 6.48 & 2.70   & $1.64\pm 0.51$\\ 
        &$10^6$&168.58 & -6.96 & 1.10 & 1.79 & 0.63  & $ 1.53\pm 0.16$ \\
        & $10^7$ & 168.02 & -6.64  & 0.35 & 0.57 & 0.32& $1.51\pm 0.05$\\\hline
        \multirow{3}{*}{5}& $10^5$ &166.00 & -6.10 & 4.34 & 7.04 & 2.11& $1.67\pm 0.51$\\
        & $10^6$ &168.70& -6.65 & 1.14 & 1.85 & 0.81 & $1.52\pm 0.16$\\
        &$10^7$ &167.95 & -6.53 & 0.37 & 0.60 & 0.41& $1.51\pm 0.06$\\\hline
        \multirow{3}{*}{10} & $10^5$ &167.01 & -5.57 & 4.60 & 7.46 & 1.65 & $1.70\pm 0.53$ \\
        &$10^6$&169.25 & -6.69 & 1.23 & 1.99 & 1.33 & $1.51\pm 0.17$\\
        &$10^7$ &168.01 & -6.49 & 0.40& 0.65 & 0.46 & $1.51\pm 0.06$\\\hline
        \multirow{3}{*}{15}&$10^5$ &166.63 & -3.19&5.13 & 8.33& 3.97 & $1.76\pm 0.59$ \\
        & $10^6$&169.45 & -6.30 & 1.25 & 2.03 & 1.63 & $1.51\pm 0.18$\\
        &$10^7$ & 167.89 & -6.66 & 0.44 & 0.72 & 0.29 &  $1.51\pm0.07$\\\hline
        \multirow{3}{*}{20} & $10^5$ &166.83 & -4.88 & 5.38 & 8.74 & 2.34& $1.84\pm 0.61$ \\
        &$10^6$&169.27 & -5.89& 1.41 & 2.29 & 1.68 & $1.53\pm 0.20$\\
        &$10^7$&167.78 & -6.76 & 0.50 & 0.81 & 0.25 & $1.52\pm 0.07$\\\hline
    \end{tabular}
    \label{tab:masksstrong}
\end{table*}
Secondly, we draw random positions on the sky and mask all cells, which centres are within $\theta=10$~deg to the random position. We start with one position and increase this masking up to 300 positions, which results in a final fractional sky coverage of $f_{sky}= 0.1$.
An example of the mask, consisting out of 100 randomly chosen positions, is also shown in Fig. \ref{fig:masks} as blue regions. We apply the masks to the simulated samples with $10^6$ sources.
The estimated dipole amplitude $\Delta d$ and the difference of the positional offset $\Delta\theta$ to the fiducial dipole as a function of sky coverage $f_{sky}$ are shown in Fig. \ref{fig:RandomCircles} and \ref{fig:RCCone}.
The estimated dipole amplitude (blue solid and red dashed line) with the estimated error (shaded region) in Fig. \ref{fig:RandomCircles} is over the complete range of sky coverage consistent with the simulated kinematic dipole (black line).
We observe a varying positional offset (black line) in Fig. \ref{fig:RCCone} from below the mean separation of the position grid to fluctuating offsets between $1.7$~deg and $5.5$~deg for the example of masking 100 random positions, depending on the simulated dipole amplitude.
The shaded regions in Fig. \ref{fig:RCCone} correspond to the $68\%$ and $95\%$ confidence cones of the 300 simulated skies. 

We conclude that the quadratic estimator (\ref{eq:estimator}) 
can reliably distinguish a kinematic dipole of order 
$10^{-3}$ from a purely random sky of at least $10^{6}$ 
point sources. Depending on the shape of the mask,
the positional offsets of the estimated dipole direction 
are of order $<3$~deg for cuts of equal galactic latitude 
or of order $5$~deg for unequal masked sky regions.

\section{Data}\label{sec:Data}
\begin{figure*}
\centering
(a)\includegraphics[width=0.32\linewidth]{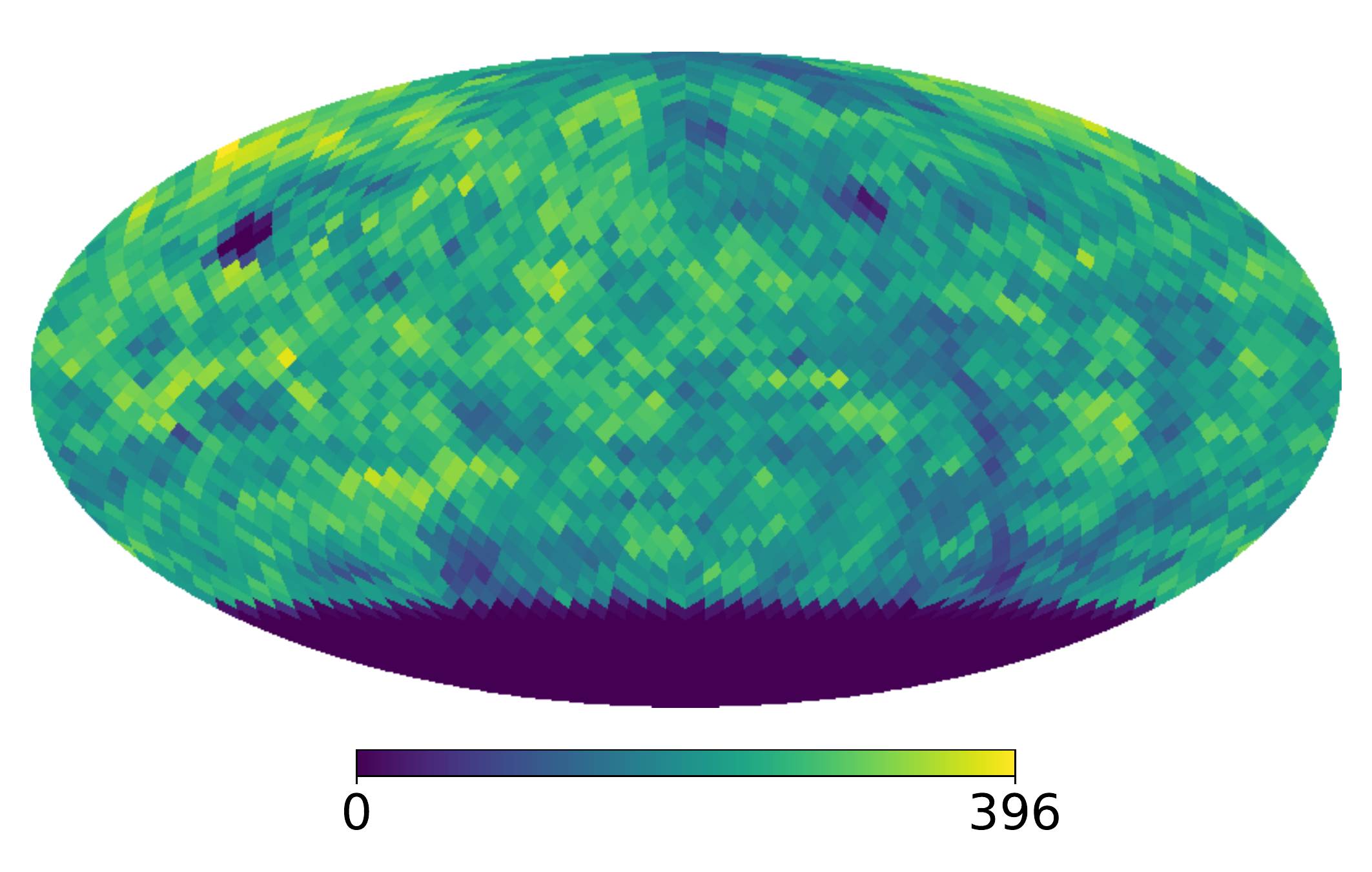}\includegraphics[width=0.32\linewidth]{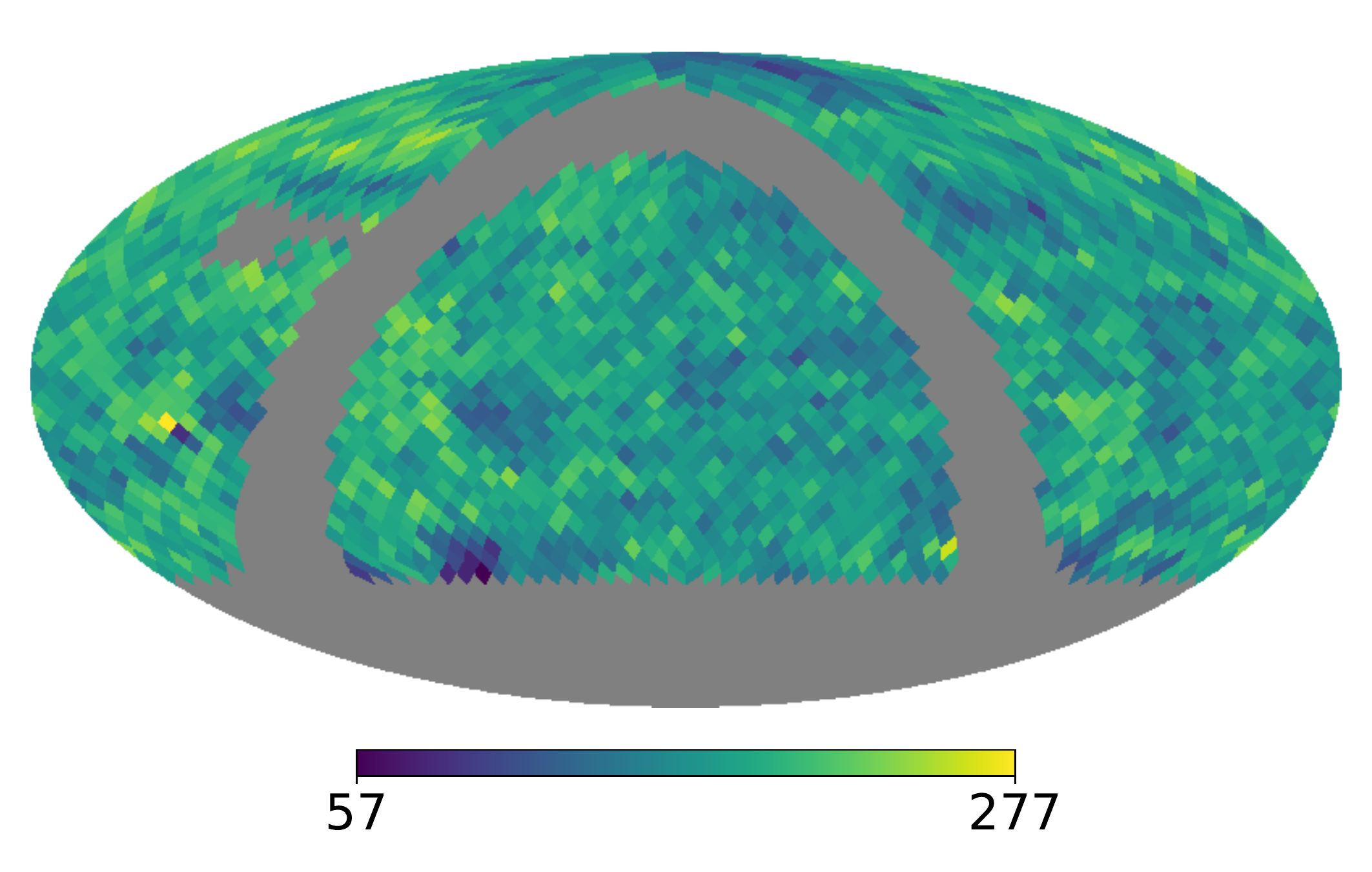}\includegraphics[width=0.32\linewidth]{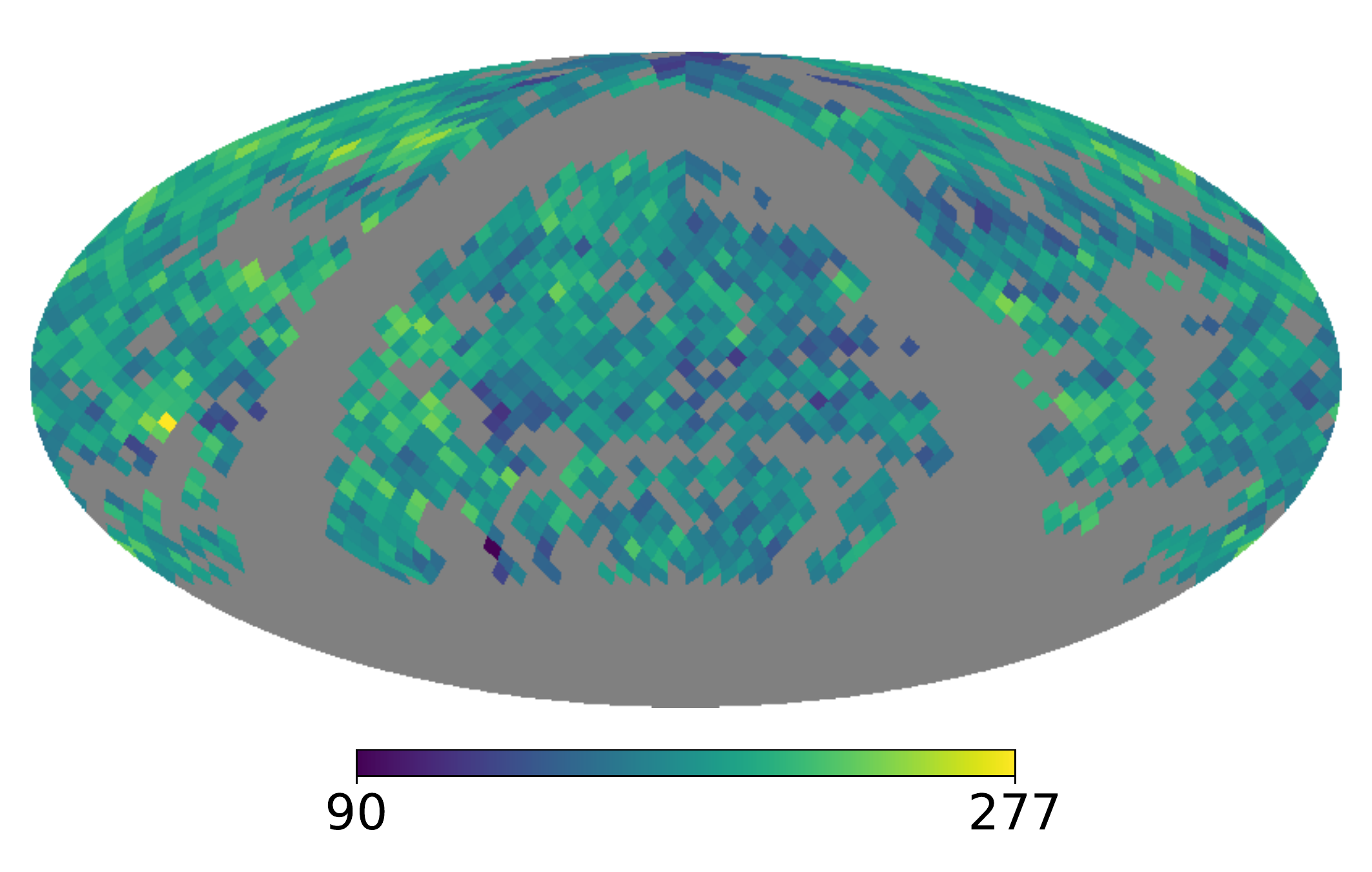}\\
(b)\includegraphics[width=0.32\linewidth]{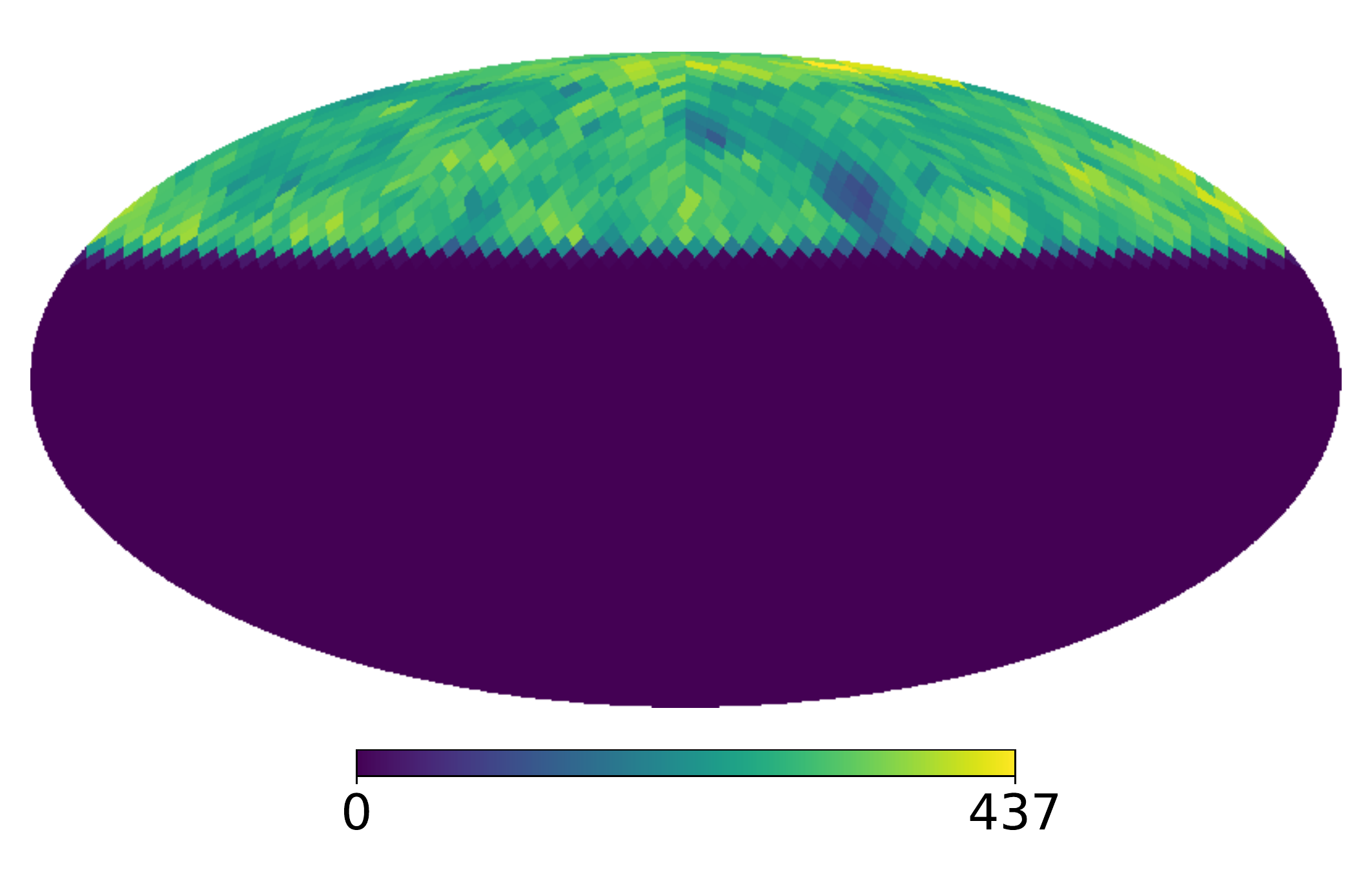}\includegraphics[width=0.32\linewidth]{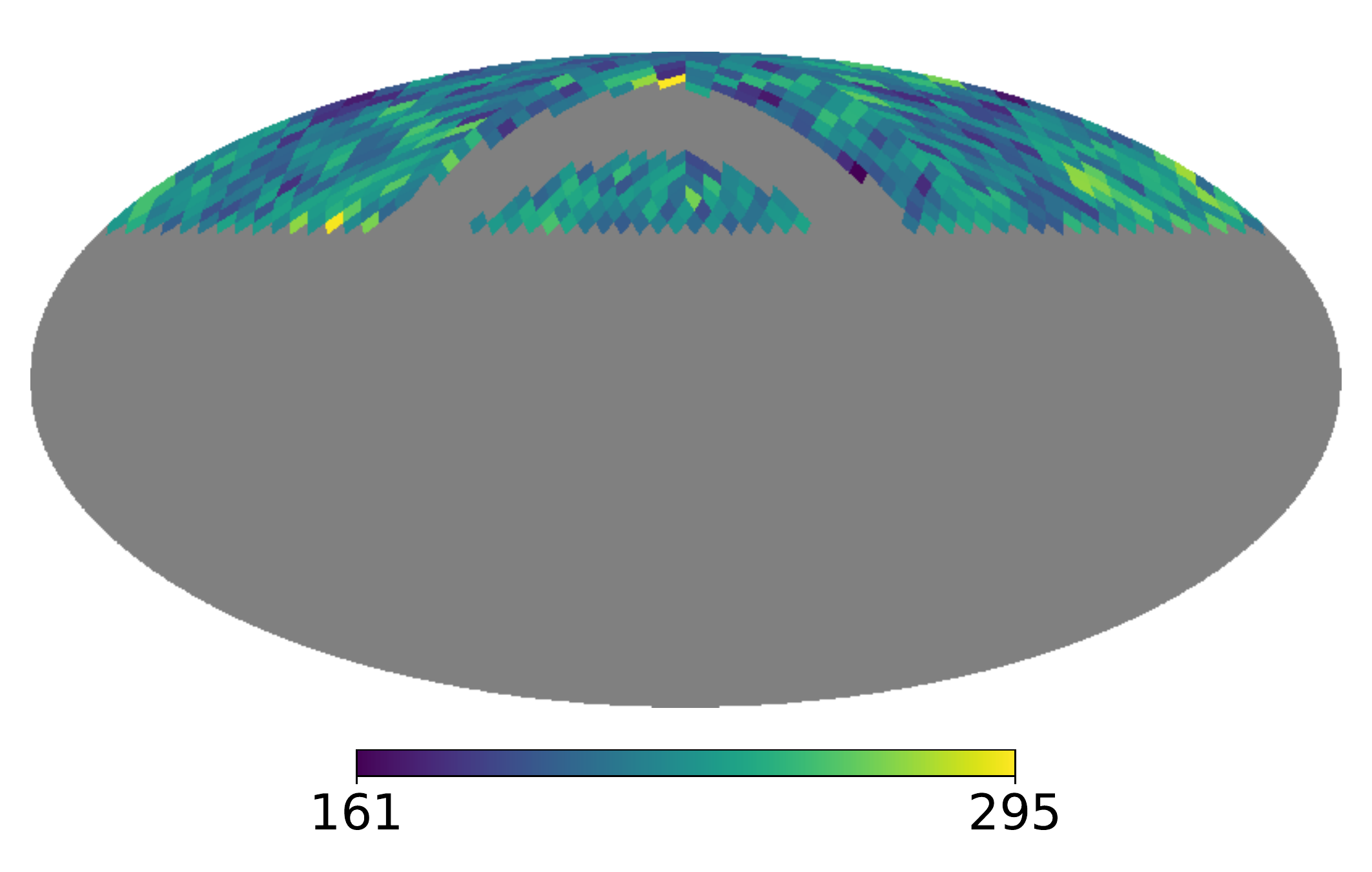}\includegraphics[width=0.32\linewidth]{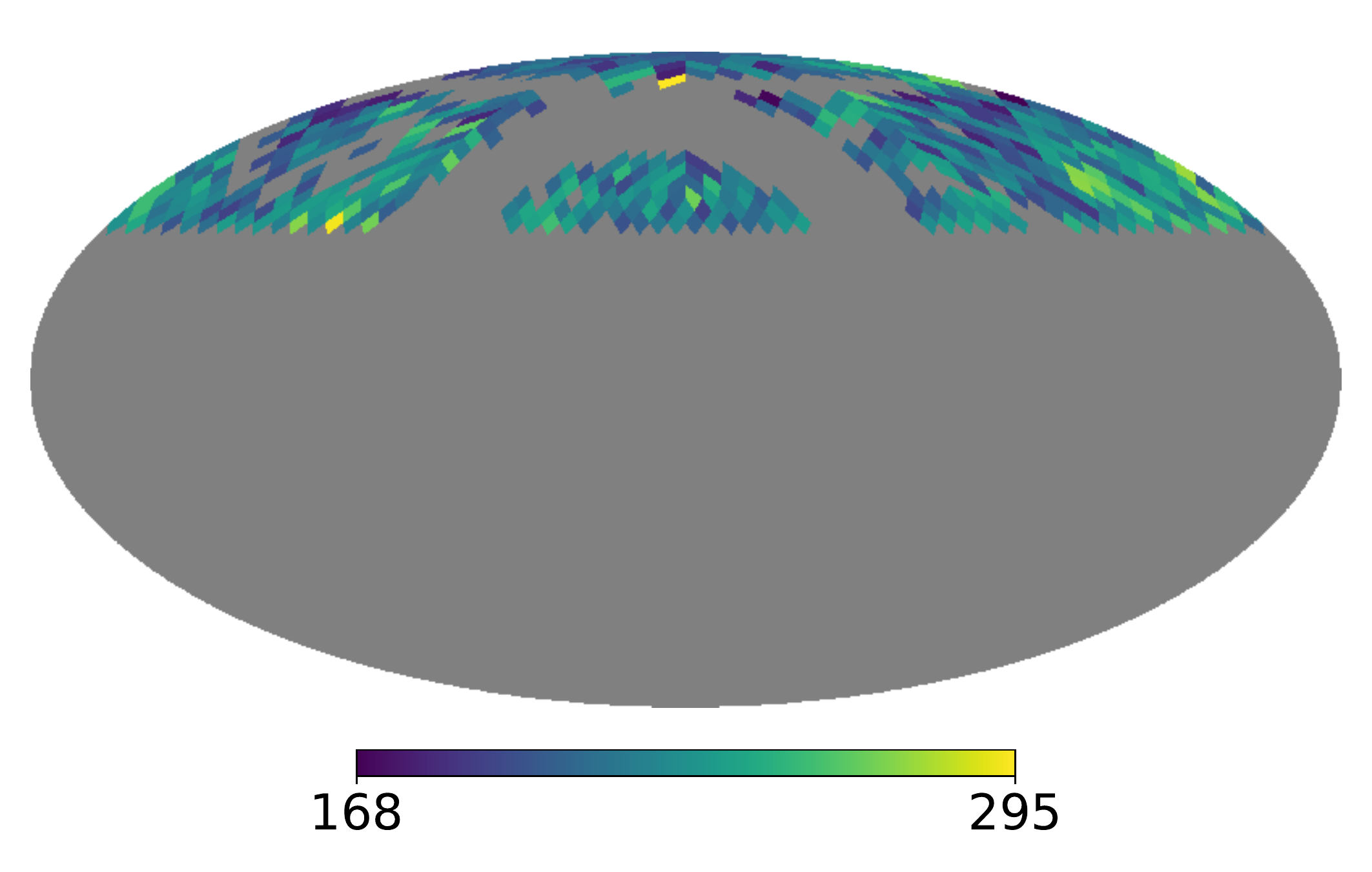}\\
(c)\includegraphics[width=0.32\linewidth]{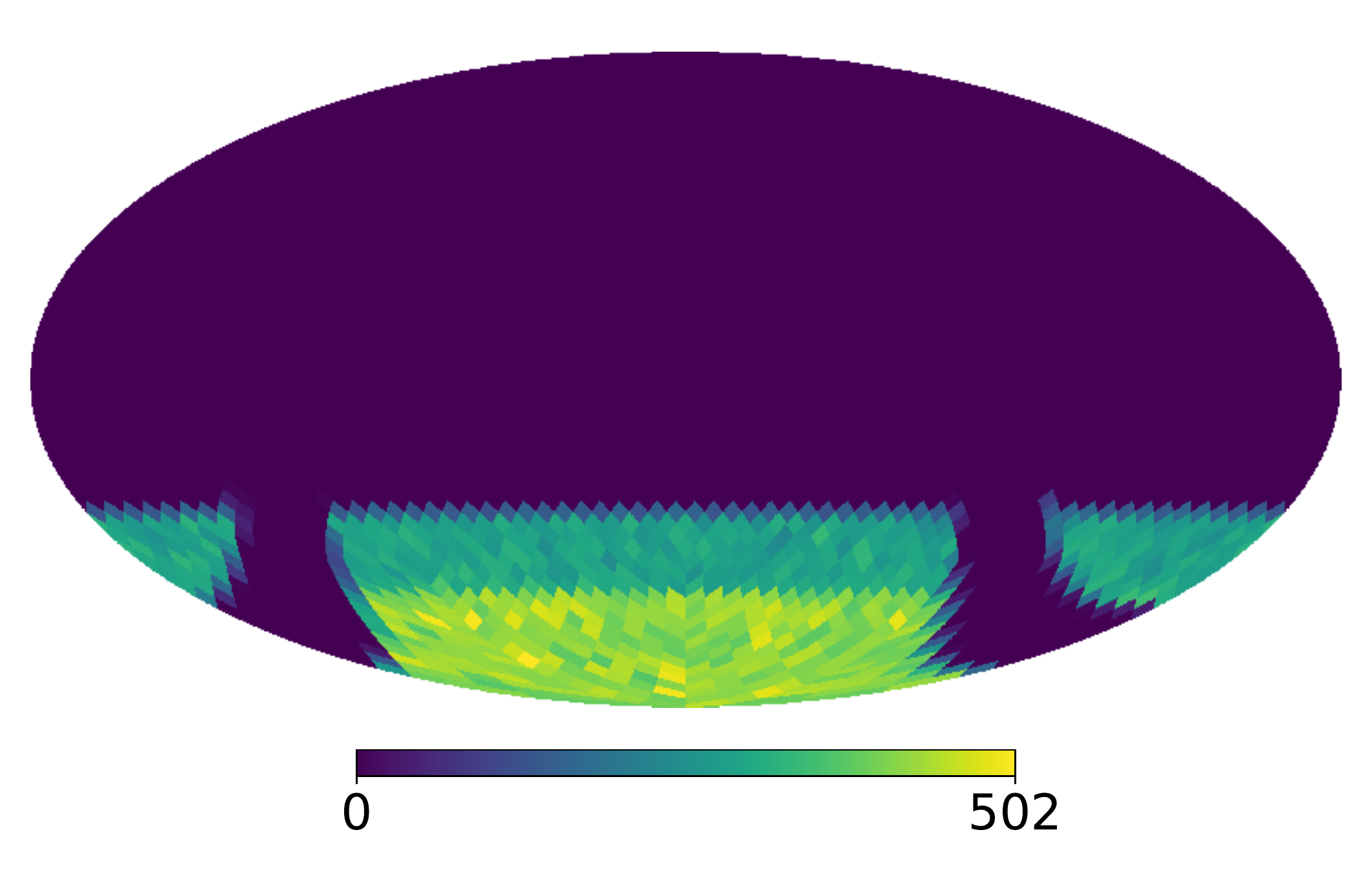}\includegraphics[width=0.32\linewidth]{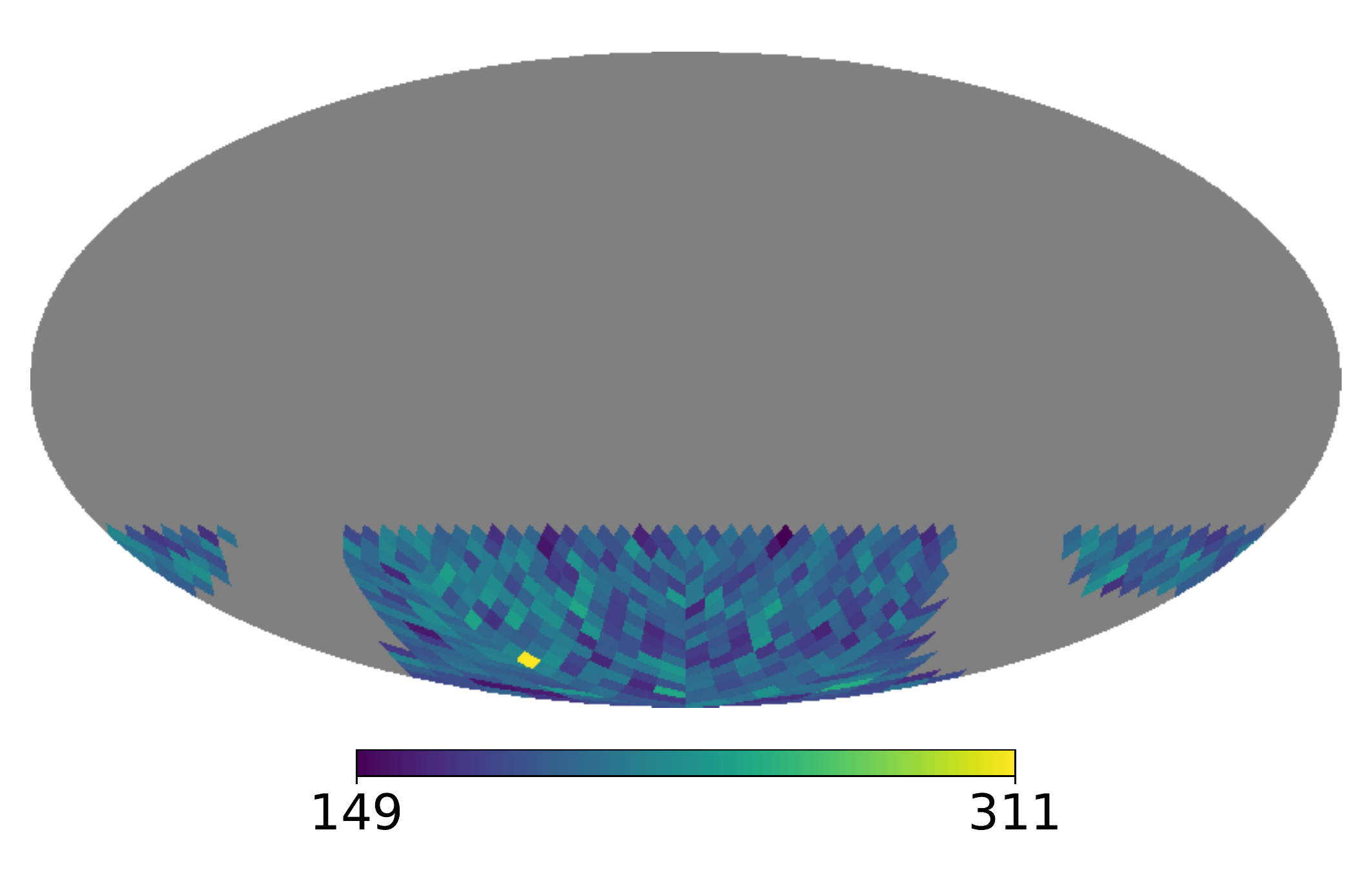}\includegraphics[width=0.32\linewidth]{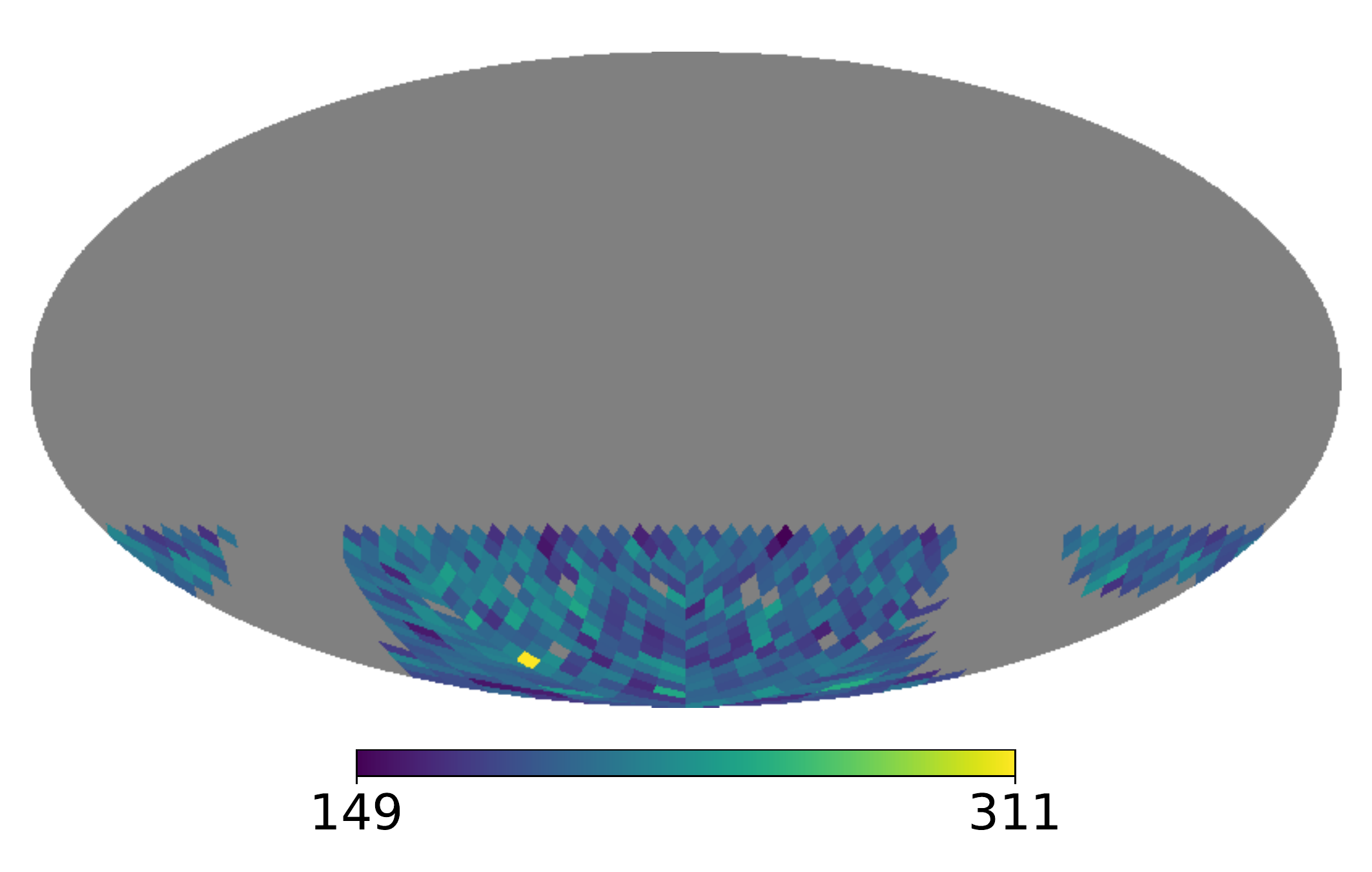}\\
(d)\includegraphics[width=0.32\linewidth]{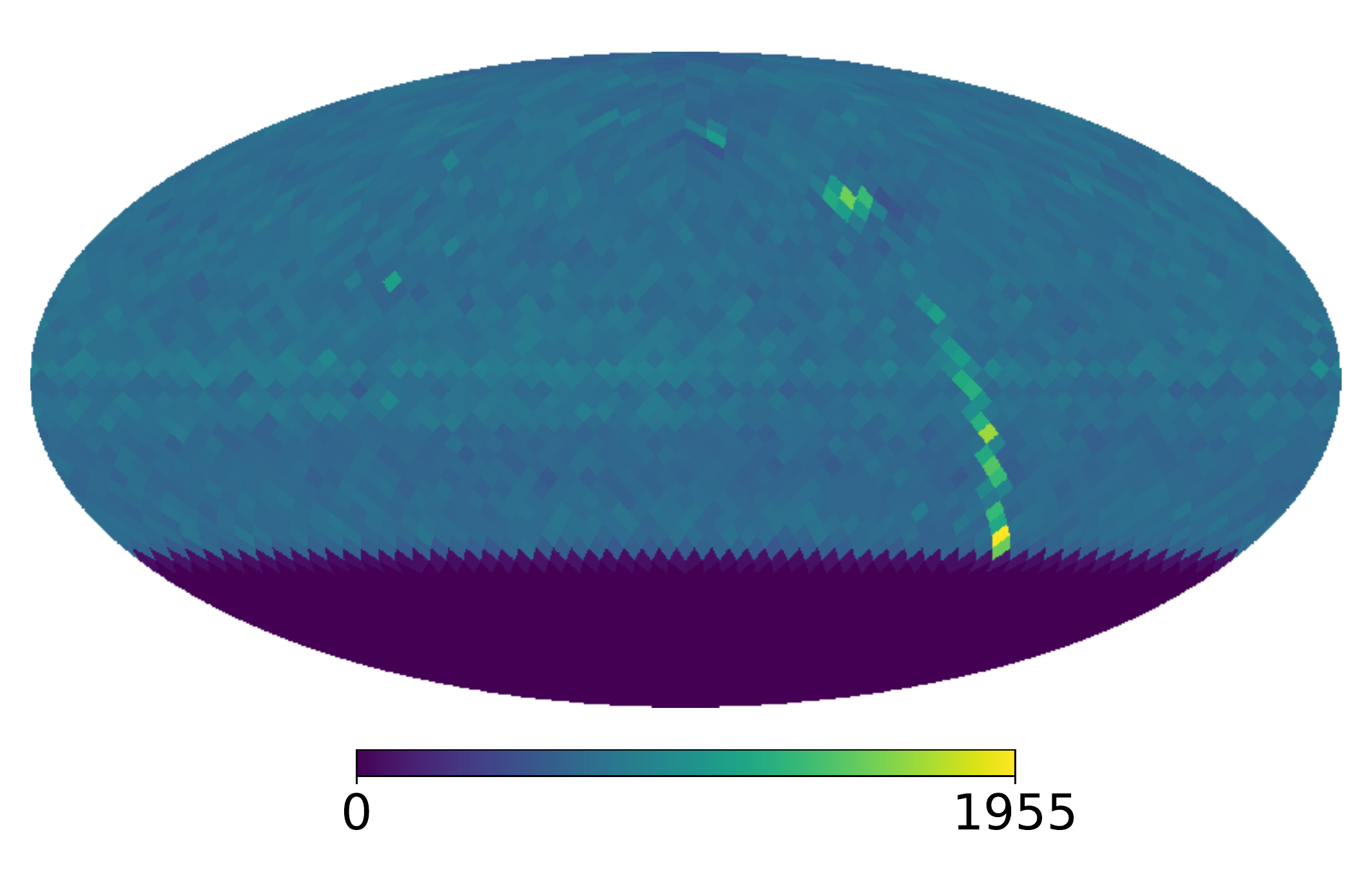}\includegraphics[width=0.32\linewidth]{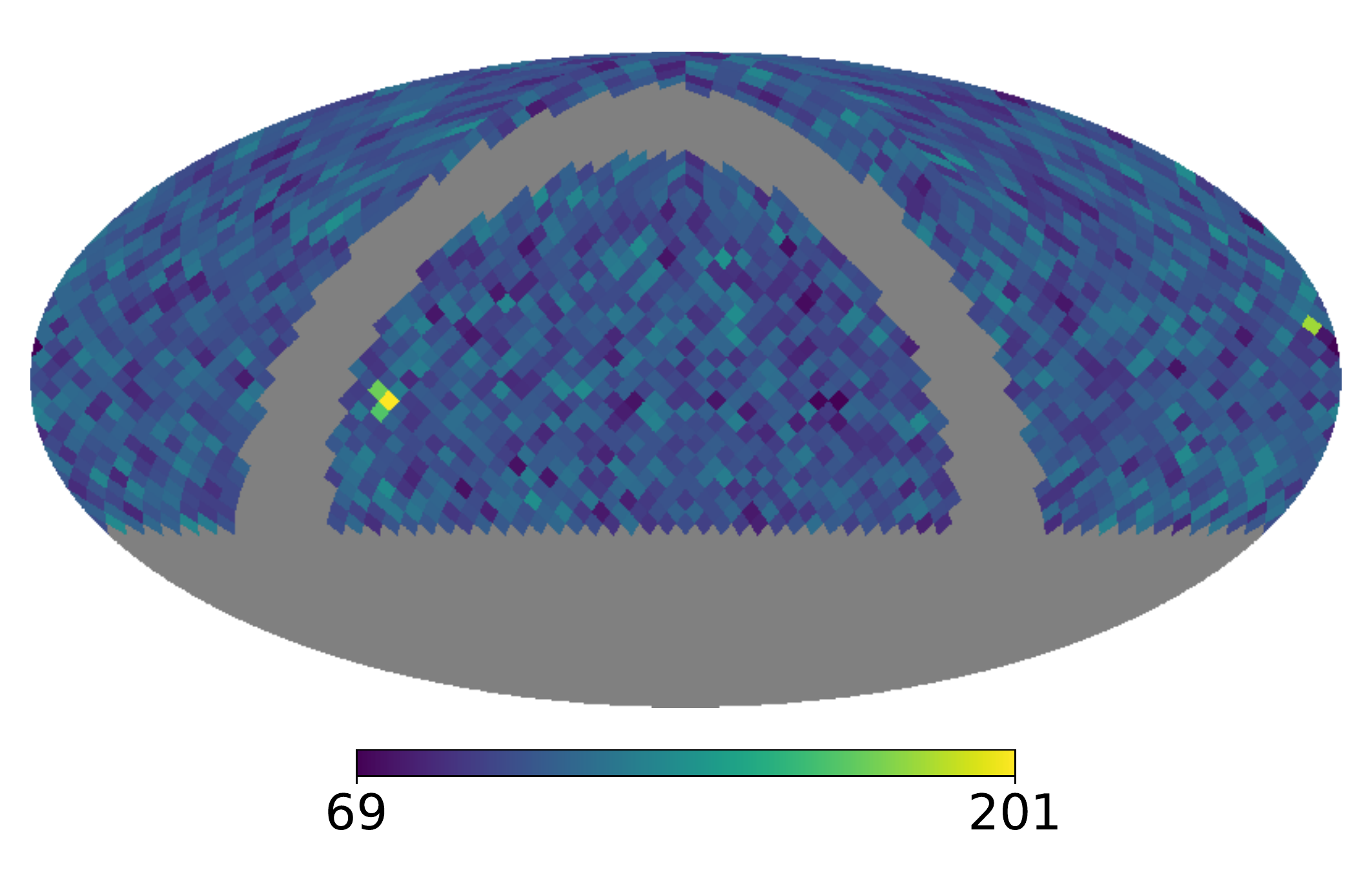}\includegraphics[width=0.32\linewidth]{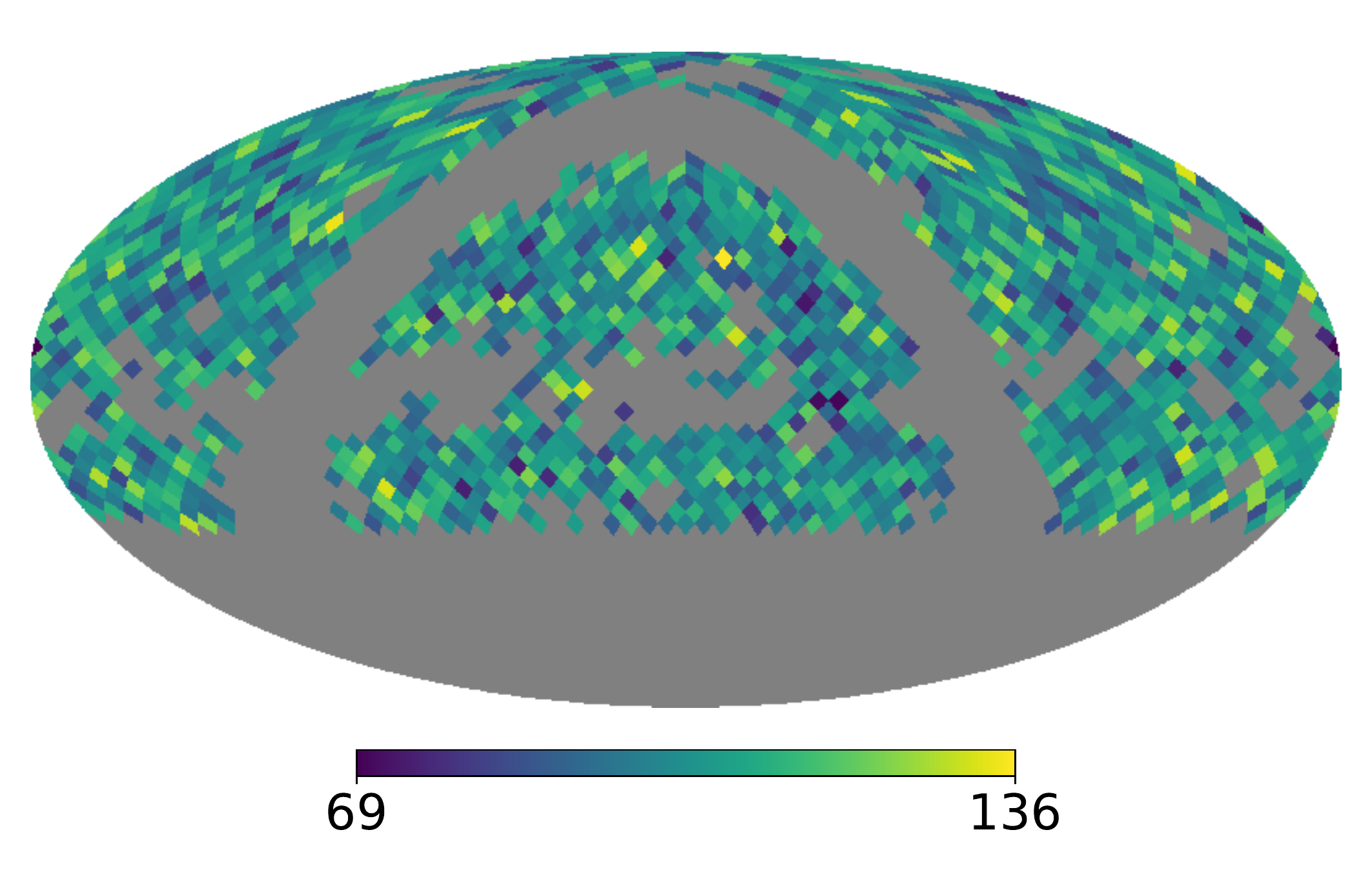}\\

\caption{Source counts per {\sc HEALPix} cell in equatorial coordinates and Mollweide
projection with {\sc HEALPix} resolution of $N_{side} = 16$. Radio sources observed within the TGSS-ADR1 (a), WENSS (b), SUMSS (c) and NVSS (d) radio source catalogues without flux density threshold are shown in the first column. In the second and third column we show the surveys masked with `mask d' and `mask n' and flux density thresholds of $50$~mJy (TGSS), $25$~mJy (WENSS), $18$~mJy (SUMSS) and $25$~mJy (NVSS).}
\label{fig:Data}
\end{figure*}

We use four different radio surveys, namely the TGSS-ADR1 \citep{TGSS2017}, WENSS \citep{WENSS1997}, SUMSS \citep{SUMSSB} and NVSS \citep{NVSS1998} to estimate the Cosmic Radio Dipole. 
The frequency range covered by these four surveys extend over one decade, starting from $147$~MHz up to $1.4$~GHz. 
We show the source counts per cell of all four surveys as a {\sc HEALPix} map with resolution of $N_{side}=16$ in Fig. \ref{fig:Data}.

\begin{table*}
  \centering
  \caption{Properties of the TGSS-ADR1, WENSS, SUMSS and NVSS. More details can be found in Sect. \ref{sec:Data}.}
  \label{tab:Data}
  \begin{tabular}{lcccc}
  \hline\hline
  & TGSS-ADR1 & WENSS & SUMSS & NVSS\\\hline
  instrument & GMRT & WRST & MOST & VLA\\
   latitude & 19$^\circ$ 05$'$ 48$''$ N & 52$^\circ$ 54$'$ 53$''$ N& 53$^\circ$ 22$'$ 15$''$ S& 34$^\circ$ 04$'$ 43$''$ N\\
  central frequency & 147. MHz &  325. MHz & 843. MHz & 1.4 GHz \\
  sky coverage& $DEC>-53$~deg &$DEC >30$~deg & $DEC\leq -30$~deg & $DEC > -40$~deg\\
  sky area & 11.24 sr & 3.10  sr & 2.47 sr& 10.32 sr\\
  sky fraction & 0.89 & 0.25 & 0.20 & 0.82\\
  limit flux density & $25$~mJy & $15$~mJy & $6/10$~mJy& $2.5$~mJy\\
  source density& $5.7\times 10^4$ sr$^{-1}$ & $7\times 10^4$~sr$^{-1}$& $8.6\times 10^4$~sr$^{-1}$& $17.2\times 10^{4}$~sr$^{-1}$\\
  resolution & $25''\times 25''$ & $54''\times 54''/\sin(DEC)$ & $45''\times45''/\cos(DEC)$ & $45''\times 45''$ \\
  number of objects& $640\,017$ &  $229\,420$  &$210\,412$ &  $1\,773\,484$ \\\hline
  \end{tabular}
  \end{table*}

\subsection{TGSS-ADR1}\label{sec:TGSS}
The first Alternative Data Release of the TIFR GMRT Sky Survey (TGSS-ADR1; \citealt{TGSS2017}) covers the sky between $-53$~deg and $+90$~deg declination, which amounts to an area of $36\,900$ square degrees, or $11.24$ sr resulting in a sky fraction of $f_{sky}\simeq 0.89$. It observes the sky with a declination dependent resolution of $25''\times 25''/\cos(DEC-19\text{ deg})$ for declinations below 19~deg and at a resolution of $25''\times 25''$ for higher declinations.
Unfortunately the observation had problems to resolve certain regions of the sky, which results into a incomplete coverage in a region roughly defined by $6^h.5\leq RA\leq 9^h.5$ and $25$~deg $\leq DEC\leq 39$~deg.
In total the survey observes $640\,017$ sources at a limiting flux density of $25$~mJy with a median rms noise of $3.5$~mJy/beam.
The survey is given to be $50\%$ complete at $25$~mJy \citep{TGSS2017} and inferred from the plotted completeness function to be $95\%$ at a flux density threshold of $100$~mJy. 
In Fig. \ref{fig:Data} the TGSS-ADR1 is shown in a {\sc HEALPix} resolution of $N_{Side}=16$, which results in a median source count per cell of $\tilde{n}=221$ and in a sample mean of $\bar{n}=222.32$ with a sample standard deviation of $\sigma = 57.89$.

\subsection{WENSS}\label{sec:WENSS}
To form the Westerbork Northern Sky Survey (WENSS) with the Westerbork Synthesis Radio Telescope (WSRT), two different catalogues have been compiled at a frequency of $325$ MHz \citep{WENSS1994}. 
These two parts are a main version, which covers the sky from $30$~deg to $75$~deg declination and a northern pole version, which contains sources above $75$~deg declination, while this part was done with a larger bandwidth to decrease the noise level.
 The resolution of the resulting mosaics is given as $54''\times 54''/\sin(DEC)$ for $325$ MHz at FWHM, while the positional accuracy for strong sources is $1.5''$. 
 
In a first mini survey of $570$~square degrees \cite{WENSS1997}, the survey is estimated to be essentially complete at $30$~mJy and by computing the differential source counts to be 50\% at 25~mJy.
The final catalogue \cite{WENSS2000}, combined out of the main and polar version, contains in total $229\,420$ sources with a limiting flux density of approximately $18\text{ mJy }$. 
It covers $10\,177$~square degrees, or $3.10$~sr of the northern sky, which compares to a fractional sky coverage of $f_{sky}=0.25$.
In Fig. \ref{fig:Data} the sky map of the WENSS survey is shown. One can see, that our own galaxy contributes in low counts to the distribution of sources in the sky.
We get a mean number of sources per cell of $\bar{n}=265.53$ and a median of $\tilde{n}=286$ with a standard deviation $\sigma_n=85.15$. 

\subsection{SUMSS}
The final version of the Sydney University Molonglo Sky Survey (SUMSS v2.0; \citealt{MGPSSUMSS}), which was observed at a central frequency of $843$~MHz, contains $210\,412$ radio sources.
The corresponding sky patch of $8100$ square degrees, or 2.47~sr has been observed for declinations $\theta\leq -30$~deg without galactic latitudes of $|b|>10$~deg, which sums up to a sky fraction of $f_{sky}=0.20$.
The resulting images have a resolution of $45''\times45''/\cos(DEC)$ with a limiting peak brightness of $6$~mJy/beam and $10$~mJy/beam, for declinations $DEC\leq-50$~deg (southern) and $DEC>-50$~deg (northern) respectively. 
In an earlier release \citep{SUMSSB} of $3500$~square degrees and $107\,765$ radio sources the two parts were estimated to be complete at $8$~mJy and $18$~mJy. 
In order to be complete in both parts of the survey, we apply the same flux density threshold to both parts and make sure it is above $18$~mJy.

The source distribution of the final survey is shown in Fig. \ref{fig:Data}, where the difference between the southern and northern part is clearly visibly by smaller source counts per cell above declinations of $-50$~deg. 
The median source count per cell of the complete sample is $\tilde{n}=303$, together with a mean of $\bar{n}=305.43$ and standard deviation of $\sigma=116.51$, which splits up into $\bar{n}_s=340.59$ and $\bar{n}_n=221.08$ for the southern and northern part, respectively. 

\subsection{NVSS}\label{sec:NVSS}
The NRAO VLA Sky Survey (NVSS; \citealt{NVSS1998}) observed at a central frequency of $1400$~MHz with a bandwidth of $42$~MHz covers $33\,813$ square degree, or $10.3$~sr of the full sky, which corresponds to a fraction of sky of about $f_{sky}= 0.82$.
Covering the sky down to $DEC=-40$~deg declination the survey contains $1\,773\,484$ sources brighter than $2$~mJy peak brightness.
The images produced out of this radio survey have a rms noise of $0.45$~mJy/beam and a resolution at FWHM of $45"$.
The source catalogue was estimated to be $50\%$ complete at a flux density threshold of $2.5$~mJy and $99\%$ complete at $3.4$~mJy \citep{NVSS1998}. 
\citet{BlakeWall2002} pointed out that the NVSS source catalogue still suffers from systematical source density fluctuations below 15~mJy.

In Fig. \ref{fig:Data} the map of the full NVSS, with a mean of source counts per cell of $\tilde{n}=691$ and a standard deviation of $\sigma=120.73$ and median of $\bar{n}=684.21$ sources with rejecting cells without sources is shown. 
One can clearly see in the map, that our own galaxy contributes in high counts to the distribution. 
The constant variation in the source counts in the region of the equatorial plane is due to a change in configurations starting at $\delta\approx-10$~deg and north of it up to $\delta\approx +75$~deg. 

\subsection{Masking}\label{sec:Masking}
All the four surveys are masked with the same strategy by applying a {\sc HEALPix} mask with $N_{Side}=16$ to the source counts per cell. 
First we cut the southern hemisphere, or northern in the case of SUMSS, to mask unobserved regions. 
The bound will be set higher, or lower to also reject cells, which centres are not below the detection limit, but overlap with their covering area. 
Secondly we mask the Galactic plane for all surveys with a cut of equal latitude $|b|\leq 10$ deg (depending on the cell centre position) in order to account for confusion by the Milky Way. 

Additionally we mask unobserved regions, for example in the case of TGSS-ADR1, where failed observations led to a large unobserved region in the north-east.
The combination out of these three masks will be used as the default mask, called `mask d'.
Based on the local rms noise we additionally generate masks defined by the cell averaged local rms noise of all sources in that given cell ($\sigma_b$) and denote it `mask n'.
We use the $68$ percentile of the rms maps as the upper bound and reject all cells with higher local noise than the $68\%$ limit. 
In case of the TGSS and WENSS survey, the local rms noise is directly given in the radio source catalogue. 
We find for the TGSS $\sigma_b^{68}=4.10$~mJy/beam and for the WENSS $\sigma_b^{68}=4.12$~mJy/beam. 
While there is no local rms noise for each source in the NVSS and SUMSS source catalogue available, we adopt the mask defined in \citet{ChenSchwarz2016}.
This mask is based on the averaged rms positional uncertainty of sources fainter than $15$~mJy in the NVSS, which traces the rms brightness fluctuations $\sigma_b$ \citep{NVSS1998}.
For the NVSS, we re-scale the mask `NVSS65' of \citet{ChenSchwarz2016} from $N_\mathrm{side}=32$ to $N_\mathrm{NSide}=16$ and named it 'mask n'.
In case of the SUMSS survey, we average the positional uncertainty per cell relative to the beam size \citep{ChenSchwarz2016}
\begin{equation}
    \sigma_\theta = \sqrt{\frac{\sigma_\phi\sigma_\theta }{\theta^2_{\mathrm{FWHM}}}\sin{\left(\frac{\pi}{2}-\theta\right)}}
\end{equation}
for all sources below the completeness level of $S=18$~mJy.
From this positional uncertainty per cell, we define the mask by only using the first $68\%$ of the cells with the limit $\sigma_\theta^{68}=6.95\times 10^{-2}$.   
The full, as well as the masked surveys can be seen in Fig. \ref{fig:Data}, with the full surveys in the left, `mask d' in the middle and `mask n' in the right column.
The reduced sky coverages of the masks are shown in Table \ref{tab:resultsdipole}.

\section{Results}\label{sec:results}  

\subsection{Expected dipole amplitude}\label{sec:expected}
Before we estimated the Cosmic Radio Dipole for all four surveys with two different masks and different flux density thresholds, we calculate the expected dipole amplitude based on the source counts of each survey. 
To each survey we apply different flux density threshold. For the TGSS we use 50, 100, 150 and 200~mJy, which are motivated by the findings of \cite{RanaBagla2019,Dolfi2019} in the terms of the angular two-point correlation function and of \cite{Singal2019} in terms of the Cosmic Radio Dipole. 
Motivated by the findings of \cite{RubartSchwarz2013} we use the same set of flux density thresholds of 25, 35, 45 and 55~mJy for WENSS, SUMSS and NVSS. For SUMSS and NVSS we extend the list of thresholds by 18~mJy and 15~mJy respectively.

\begin{figure}
\centering
\includegraphics[width=\linewidth]{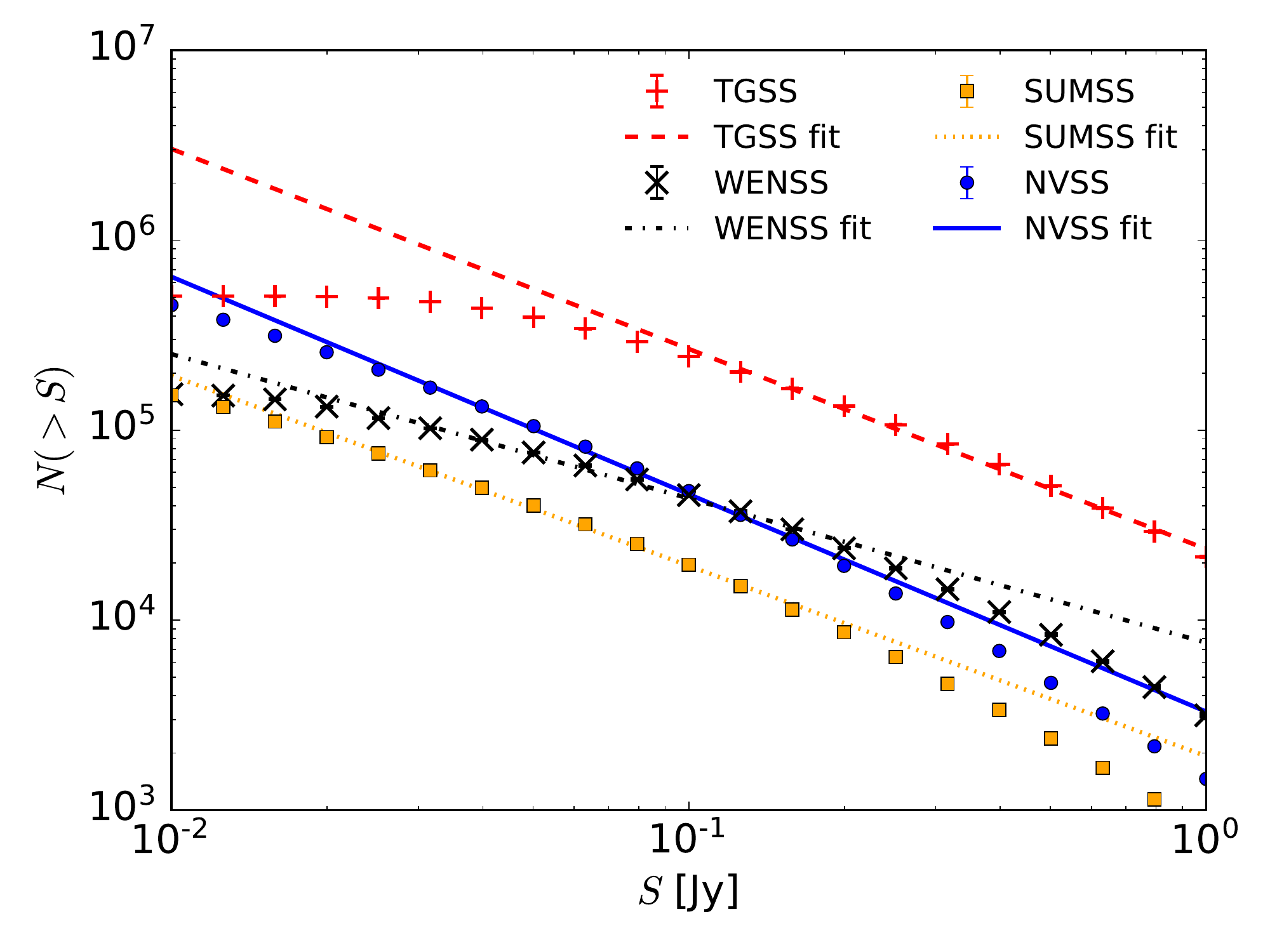}\\
\includegraphics[width = \linewidth]{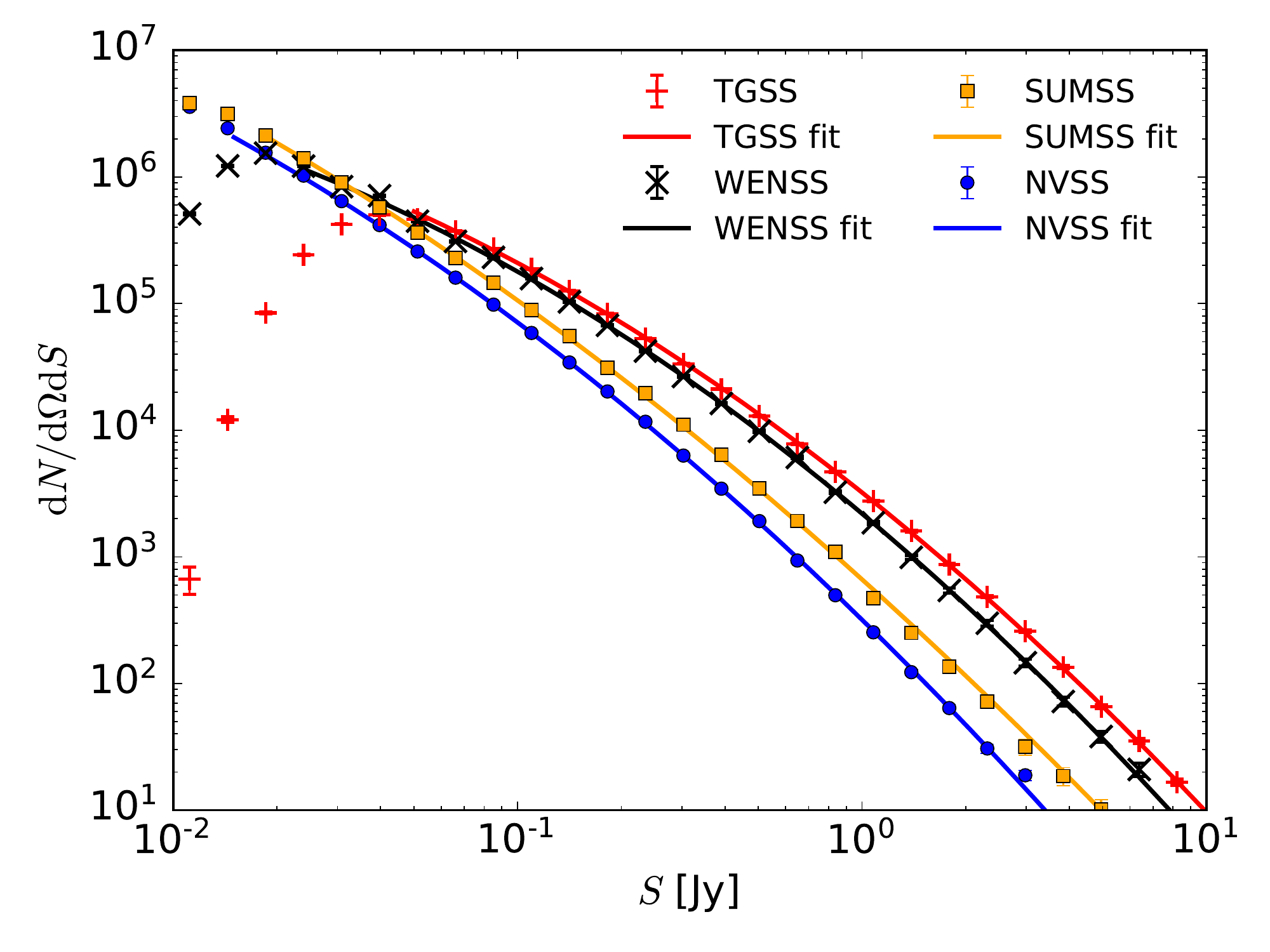}
\caption{Integral source counts of the TGSS-ADR1, WENSS, SUMSS and NVSS radio source catalogues, masked with `mask d', as a function of flux density. We fit a function $N(>S)\propto S^{-x}$ (top) and a more sophisticated function $\mathrm{d}N/\mathrm{d}\Omega\mathrm{d}S\propto xS^{-x-\kappa\log_{10}(S/\text{Jy})}$ (bottom) to the counts, which have equal step width in $\log_{10}(S/\text{Jy})$. The fitted range depends on the flux threshold of the survey and extends for each survey over one decade.}
\label{fig:comparisonlogfit}
\end{figure}

As denoted in Eq. (\ref{eq:sourcedistribution}), the source counts per solid angle and flux density threshold can be defined as a simple power law. 
In order to fit this power law, we use the least-square method of {\sc lmfit}.
The results of the fit to the unmasked source counts of the four source catalogues can be seen in Fig. \ref{fig:comparisonlogfit} (top).
Within $68\%$ confidence intervals we find for the unmasked surveys:
\begin{equation}
\begin{aligned}
&x_{\mathrm{TGSS}}=0.86\pm 0.03, &50\text{ mJy }\leq S\leq500\text{ mJy }\\
&x_{\mathrm{WENSS}}=0.75\pm0.03, &25\text{ mJy }\leq S\leq250\text{ mJy }\\
&x_{\mathrm{SUMSS}}=1.00\pm 0.02, &18\text{ mJy }\leq S\leq180\text{ mJy }\\
&x_{\mathrm{NVSS}}=1.04\pm 0.02, &15\text{ mJy }\leq S\leq150\text{ mJy }\\
\end{aligned}
\end{equation}
The lower bound for the fitting range is defined by the smallest flux density threshold that we apply to each survey. The fitting range is then extended over one decade of flux density.
We additionally perform the same fit as above, but using the surveys masked with `mask d' and `mask n'.
\begin{equation}
\begin{aligned}
&\mathrm{mask\,d}\\
&x_{\mathrm{TGSS}}=0.87\pm 0.03\\
&x_{\mathrm{WENSS}}=0.76\pm 0.03\\
&x_{\mathrm{SUMSS}}=1.00\pm 0.02\\
&x_{\mathrm{NVSS}}=1.04\pm 0.02\\
\end{aligned}\qquad
\begin{aligned}
&\mathrm{mask\,n}\\
&x_{\mathrm{TGSS}}=0.88\pm 0.03\\
&x_{\mathrm{WENSS}}=0.77\pm 0.03\\
&x_{\mathrm{SUMSS}}=1.00\pm 0.02\\
&x_{\mathrm{NVSS}}=1.05\pm 0.02\\
\end{aligned}
\end{equation}

The fitted slopes of the source counts are consistent between the unmasked and masked surveys, but tend to steepen within error bars with more aggressive masks. 
These findings are in good agreement with earlier studies, e.g. \cite{RubartSchwarz2013}.
For NVSS and SUMSS the results are additionally in good agreement to the usual assumption of $x\approx1$.

Based on the results in Fig. \ref{fig:comparisonlogfit} (top), the assumption of a single power law seems unlikely.
\citet{Jain13} suggested a deviation from a pure power law, which extends the slope by an extra dependency on the flux density.
The improved power law is then defined on the differential source counts as:
\begin{equation}\label{eq:improvedpowerlaw}
    \frac{\mathrm{d}N}{\mathrm{d}\Omega\mathrm{d}S}\propto xS^{-1-x-\kappa \ln(S/\text{Jy})}.
\end{equation}
The fit of Eq. (\ref{eq:improvedpowerlaw}) is carried out in the range from the lowest defined flux density threshold up to 10~Jy.
In Fig. \ref{fig:comparisonlogfit} (bottom) we show the results of this fit to the differential source of the masked surveys.
The resulting best-fit values for $x$ and $\kappa$ within $68\%$ confidence intervals are:
\begin{equation}
\begin{aligned}
&x_{\mathrm{TGSS}}=1.168\pm 0.003, \\
&x_{\mathrm{WENSS}}=1.286\pm0.032,  \\
&x_{\mathrm{SUMSS}}=1.453\pm 0.022, \\
&x_{\mathrm{NVSS}}=1.662\pm 0.014,  \\
\end{aligned}\quad 
\begin{aligned}
&\kappa_{\mathrm{TGSS}} = 0.155\pm 0.002\\
&\kappa_{\mathrm{WENSS}} = 0.163\pm 0.008\\
&\kappa_{\mathrm{SUMSS}} = 0.108\pm 0.004 \\
&\kappa_{\mathrm{NVSS}} = 0.136\pm 0.003\\
\end{aligned}
\end{equation}
Based on the improved power law of Eq. (\ref{eq:improvedpowerlaw}), the expected dipole amplitude can be written as \citep{Jain13}:
\begin{equation}\label{eq:improveddipoleamplitude}
    d_i = \left[2+x(1+\alpha)+2\kappa(1+\alpha)\frac{I_2}{I_1}\right]\beta
\end{equation}
with:
\begin{align}
    I_1 &= x\int_{S_{lim}}^{10}\mathrm{d}S~S^{-1-x-\kappa \ln(S/\text{Jy})},\\
    I_2 &= x\int_{S_{lim}}^{10}\mathrm{d}S~S^{-1-x-\kappa \ln(S/\text{Jy})}\ln(S/\text{Jy}).
\end{align}
In Table \ref{tab:expecteddipole}, we compare the expected dipole amplitudes from Eq. (\ref{eq:dipoleamplitude}) and Eq. (\ref{eq:improveddipoleamplitude}).
The errors of the best-fit values are propagated to estimate the error of the expected dipole amplitudes.
Additionally we assume the error of the spectral index to be $\Delta\alpha=0.25$ and propagate it accordingly.
For WENSS, SUMSS and NVSS the expected dipole calculated from the simple fit gives stronger amplitudes than for the improved fit. 
In the case of TGSS the usual fit gives a significant weaker dipole amplitude compared to the improved fit.
The resulting expectations of the simple fit for WENSS and NVSS are also in good agreement to expectations of \citet{RubartSchwarz2013}, with $d_{WENSS}=(0.42\pm 0.03)\times10^{-2}$ and $d_{NVSS}=(0.48\pm 0.04)\times10^{-2}$ respectively. 

\begin{table}
    \centering
    \caption{Expected dipole amplitudes for the TGSS, WENSS, SUMSS and NVSS radio source catalogues, masked with `mask d'. The dipole amplitudes $d$ and $d_i$ are computed from Eq. (\ref{eq:dipoleamplitude}) and (\ref{eq:improveddipoleamplitude}) respectively. }
    \begin{tabular}{ccc}
        \hline\hline
        Survey & $d$ & $d_i$  \\
        & $(\times 10^{-2})$ & $(\times 10^{-2})$\\\hline
         TGSS & $0.434\pm 0.027$ & $0.500\pm0.036$ \\
         WENSS &$0.410\pm 0.024$ & $0.357\pm0.028$\\
         SUMSS &$0.463\pm 0.031$& $0.420\pm0.035$\\
         NVSS & $0.472\pm 0.033$ & $ 0.418\pm 0.038$\\\hline
    \end{tabular}
    \label{tab:expecteddipole}
\end{table}

\subsection{Measurements of the cosmic radio dipole}\label{sec:subresults}

The results of the estimation of the Cosmic Radio Dipole from the four different surveys are shown in Table \ref{tab:resultsdipole}. 
From the estimated dipole direction $(\phi_m,\theta_m)$ we calculated the angular distance $\Delta \theta$ to the fiducial kinematic dipole $(\phi_f,\theta_f)$ of the CMB (see Sect. \ref{sec:cosmicradiodipole}) using Eq. (\ref{eq:offset}).
The errors of each estimation are computed from 100 bootstraps of the source counts map. 
Additionally we add half the cell size of the position grid to the estimated error in order to account for the discretization. 
In our case we use a position grid of $N_{side}=32$ resolution, so cells have a mean spacing of $\Delta\theta=1.83$~deg.
The directional offset error was computed by taking the standard deviation of the directional offsets from the dipole boosted skies of the surveys:
\begin{equation}
\sigma_{\Delta\theta}= \sqrt{\frac{1}{N}\sum_{i}^{N}\left(\Delta\theta_i-\Delta\theta_{\mathrm{obs}}\right)^2}
\end{equation}

We observe consistent dipole directions and dipole amplitudes for each survey between the two different masking strategies. 
We find good self-consistency for the dipole amplitude between the estimates from WENSS and SUMSS.
For the NVSS we find smaller dipole amplitudes than for WENSS and SUMSS, but they are consistent within one to two sigma. 
The estimated dipole amplitude for the TGSS shows a significant excess by a factor of two to three while comparing to the other surveys. 
In general all four surveys show an increased dipole amplitude, as well as a directional offset in comparison to the kinematic CMB dipole.
Based on the $\chi^2/dof$, we find the best agreement of the dipole modulation to the observed source counts for the NVSS catalogue.
Additionally, the smallest amplitude is found for NVSS with a flux density threshold of $15$~mJy and masked with `mask n'. 
Whereas the smallest directional offset to the CMB dipole of 1.54~deg is found for NVSS with 45~mJy and `mask n'.
The estimated dipole direction of SUMSS shows the largest offsets from the kinematic CMB dipole direction, which is mainly driven by the values in right ascension. 
The estimated declination is still in good agreement with the CMB.
In most cases the uncertainties of the dipole direction directly increase with decreasing number of sources and decreasing dipole amplitude.
The former was already expected from the testing in Sect. \ref{sec:quadraticestimator}.
Driven by the large dipole amplitude found for the TGSS we observe relatively small errors for the estimated right ascension and declination as compared to the results from the other surveys.
The results of the Cosmic Radio Dipole found in the TGSS, WENSS and NVSS are in good agreement to results from the literature.
A detailed list of earlier results from several studies are presented in App. \ref{sec:otherresults}.

\begin{table*}
    \centering
    \caption{Estimated directions ($RA$,$DEC$) and amplitudes ($d$) of the Cosmic Radio Dipole for a given survey with masks and flux density thresholds ($S$). We also show the resulting number of sources ($N$) and the directional offset ($\Delta\theta$) to the fiducial kinematic dipole of the CMB, as well as $\chi^2/dof$ of the fit. Errors are estimated from 100 bootstraps for each sample. To ease comparison across frequencies we also include scaled flux density thresholds, at a reference frequency of $1.4$~GHz ($S_{1.4}$), assuming a universal power-law spectrum with $\alpha = 0.75$.}
    \begin{tabular}{ccccccccccc}
        \hline\hline
        Survey & Mask & $f_{sky}$ & $S$ & $S_{1.4}$ & $N$ & $RA$ & $DEC$ & $\Delta\theta$ & $d$ & $\chi^2/dof$ \\
         & & &(mJy) & (mJy) & & (deg) & (deg) & (deg) & $(\times 10^{-2})$ & \\\hline
         \addlinespace[1ex]\multirow{8}{*}{TGSS} &  \multirow{4}{*}{d} &  \multirow{4}{*}{0.72} & 50 &9& $393\,447$ & $124.53\pm 4.13$ & $25.66\pm 5.15$ & $53.30\pm 4.02$ &$6.6\pm0.5$ & 3.19 \\
         &&&100&18&$244\,881$& $135.61\pm 11.57$ & $15.90\pm 11.24$ & $39.33\pm 14.30$ & $6.0\pm 0.8$ & 2.91 \\
         &&&150&28&$173\,964$& $139.53\pm 11.33$  & $12.88\pm 10.74$ & $34.50\pm 13.86$ &$5.9\pm 0.7$ & 1.83 \\
         &&&200&37&$133\,547$ & $141.99\pm 11.17$ & $11.52\pm 10.21$ & $31.74\pm 13.29$ & $5.9\pm 0.7$ & 1.65 \\
         \addlinespace[1ex]&  \multirow{4}{*}{n}  &  \multirow{4}{*}{0.52} & 50 &9&$296\,855$ &  $132.90\pm 4.57$ & $15.68\pm 5.21$ & $41.43\pm 4.17$& $6.2\pm 0.5$ & 2.36 \\
         &&&100&18&$179\,951$&$137.25\pm 6.62$ & $14.49\pm 5.39$ & $37.23\pm 6.05$ & $ 6.3\pm 0.6$&1.94 \\
         &&&150&28&$127\,244$ & $138.30\pm 6.25$ & $14.96\pm 5.25$ & $36.65\pm 5.63$& $6.5\pm 0.7$ &1.72 \\
         &&&200&37&$97\,355$& $138.86\pm6.12$ & $15.79\pm 5.51$ & $36.69\pm 5.45$ & $6.8\pm 0.8$ &1.54 \\\addlinespace[1ex]\hline
         \addlinespace[1ex]\multirow{8}{*}{WENSS} &   \multirow{4}{*}{d}  &  \multirow{4}{*}{0.17}& 25 &8& $115\,808$ &  $143.34\pm 19.48$ & $-13.15\pm 4.58$ & $24.99\pm 13.84$ & $3.2\pm 1.0$ & 1.91 \\
         &&&35&12&$95\,302$& $137.85\pm 24.47$ & $-13.29\pm 4.98$ & $30.27\pm 18.99$ & $2.9\pm 0.9$ &1.77 \\
         &&&45&15&$81\,534$ & $131.83\pm 27.76$ & $-11.95\pm 6.28$ & $35.94\pm 22.94$ & $2.8\pm 0.9$ & 1.68 \\
         &&&55&18&$71\,643$& $127.51\pm 29.27$ & $-10.70\pm 6.59$ & $40.10\pm 24.89$ & $2.8\pm 0.9$ & 1.57 \\
         \addlinespace[1ex]&  \multirow{4}{*}{n}  &  \multirow{4}{*}{0.14}& 25 &8& $93\,577$ &  $142.20\pm 23.25$ & $-16.20\pm 5.77$ & $26.83\pm 14.94$ & $3.1\pm 0.9$ & 1.88 \\
         &&&35&12&$76\,760$& $138.98\pm 27.58$ & $-16.25\pm 6.16$ & $29.81\pm 18.54$ & $2.9\pm 0.9$ & 1.75 \\
         &&&45&15&$65\,494$& $138.71\pm 34.24$ & $-16.23\pm 7.66$ & $30.06\pm 23.10$ & $2.8\pm 1.0$ & 1.67 \\
         &&&55&18&$57\,463$&$135.43\pm 35.16$& $-15.39\pm 7.60$ & $32.95\pm 24.13$& $2.8\pm 1.0$ & 1.56 \\\addlinespace[1ex]\hline
         \addlinespace[1ex]\multirow{10}{*}{SUMSS} &  \multirow{5}{*}{d}  &  \multirow{5}{*}{0.16}& 18  &12& $99\,835$ &  $106.67\pm12.90$  & $-9.50\pm 11.12$ & $60.62\pm 12.49$ & $3.8\pm 0.9$ & 1.49 \\
         &&&25&17&$75\,642$ & $106.18\pm 16.99$ & $-5.11\pm 9.91$ & $61.40\pm 16.79$ & $3.5\pm 1.0$ & 1.58 \\
         &&&35&24&$55\,973$& $108.05\pm 22.64$ & $-4.12\pm 8.92$ & $59.65\pm 20.85$ & $3.4\pm 1.0$ & 1.49 \\
         &&&45&31&$44\,403$& $105.33\pm 25.64$ & $-4.08\pm 8.35$ & $62.35\pm 23.73$ & $3.3\pm 1.1$ & 1.51 \\
         &&&55&38&$36\,646$& $106.72\pm 33.92$ & $-4.92\pm 8.66$ & $60.89\pm 27.50$ & $3.2\pm 1.1$ & 1.40 \\
        \addlinespace[1ex] &  \multirow{5}{*}{n}&  \multirow{5}{*}{0.16} & 18  &12& $96\,816$ &  $106.67\pm 14.53$ & $-9.50\pm 10.03$ & $59.40\pm 14.36$ & $3.8\pm 0.8$ & 1.51 \\
         &&&25&17&$73\,356$& $106.18\pm 17.34$ & $-5.11\pm 8.95$ & $61.16\pm 17.28$ & $3.5\pm 1.0$ & 1.60 \\
         &&&35&24&$54\,336$& $108.05\pm 20.78$ & $-4.12\pm 8.16$ & $61.24\pm 20.09$ & $3.4\pm 1.1$ & 1.51 \\
         &&&45&31&$43\,121$& $105.33\pm 24.68$ & $-4.08\pm 7.93$ & $63.50\pm 23.62$ & $3.3\pm 1.1$ & 1.46 \\
         &&&55&38&$35\,574$& $106.72\pm 30.58$ & $-4.92\pm 8.68$ & $61.60\pm 25.75$ & $3.2\pm 1.2$ &1.41 \\\addlinespace[1ex]\hline
        \addlinespace[1ex] \multirow{10}{*}{NVSS} &  \multirow{5}{*}{d}&  \multirow{5}{*}{0.66} & 15&--&  $328\,207$ & $138.90\pm 12.02$ & $-2.74\pm 12.11$ & $29.23\pm 11.07$ & $1.6\pm 0.3$ & 1.30 \\
         &&& 25  &--& $209\,034$ & $140.02\pm 13.63$  & $-5.14\pm 13.26$ & $27.82\pm 12.17$ & $1.8\pm 0.4$ & 1.23 \\
         &&&35&--&$151\,702$ & $140.51\pm14.14$ & $-8.32\pm 14.52$ & $27.22\pm 12.61$ & $1.8\pm 0.4$ & 1.23 \\
         &&&45&--&$117\,617$& $140.67\pm 14.68$ & $-13.01\pm 16.15$ & $27.52\pm 12.65$ & $2.0\pm 0.6$ &1.24 \\
         &&&55&--&$95\,129$ & $143.86\pm 17.03$ & $-16.45\pm 17.38$ & $25.39\pm 12.76$ & $2.1\pm 0.6$ & 1.23 \\
          \addlinespace[1ex]&  \multirow{5}{*}{n} &  \multirow{5}{*}{0.53} & 15 &--& $266\,839$ & $ 156.33\pm 17.80$&$7.41\pm 17.63$ & $ 18.44\pm 15.16$ & $1.4\pm 0.4$ & 1.18\\
          &&& 25 &--& $169\,752$ & $161.02\pm 17.37$ & $2.69\pm 17.12$ & $ 11.86\pm 13.94$ & $1.6\pm 0.4$ & 1.10\\
          &&& 35 &--& $123\,037$ & $165.14\pm 18.88$ &	$-1.84\pm 18.82$ & $5.82\pm 13.65$ & $1.6\pm 0.5$ & 1.13\\
          &&& 45 &--& $95\,291$ & $169.15\pm 19.40$ & $-5.99\pm 19.29$ & $1.54\pm 13.05$ & $1.8\pm 0.5$ & 1.10\\
          &&& 55 &--& $77\,081$ & $173.60\pm 21.09$ & $-9.18\pm 19.47$ & $6.03\pm 13.47$ & $2.0\pm 0.6$ & 1.10\\\addlinespace[1ex]\hline
    \end{tabular}
    \label{tab:resultsdipole}
\end{table*}

The excess of the dipole amplitude found in the TGSS, as compared to other surveys, is within two sigma agreement to results of \citet{Bengaly2018} and \citet{Singal2019} (see Tab. \ref{tab:resultstgss}).
As the TGSS is reported to be incomplete below $100$~mJy the smallest flux density threshold in our TGSS samples shows the largest offset to the CMB dipole direction.
Within the `mask d' samples the incomplete sample additionally shows the largest dipole amplitude.
The three other flux density threshold samples show a self-consistent dipole amplitude and dipole direction.
The `mask n' results show a slightly increased directional offset and slightly increased dipole amplitude, but are self-consistent.
Based on the $\chi^2/dof$, the agreement of the dipole hypothesis to the data improves continuously for increasing flux density thresholds.
The faintest TGSS samples of both masks show a significantly lager $\chi^2/dof$ than higher flux density thresholds.

In general all our measured amplitudes of the WENSS `mask d' and `mask n' samples are consistent within one sigma.
For the faintest samples however we find the largest dipole amplitude, which decreases slightly for higher flux density thresholds.
The results of our WENSS samples are also in good agreement to the results reported in  \citet{RubartSchwarz2013} (see. Table \ref{tab:resultswenss}), which used a linear estimator.
The increasing offset of the dipole direction in the WENSS samples is mostly driven by the measured right ascension angles.
We observe a decreasing right ascension for higher flux density thresholds and a with the CMB dipole direction consistent declination.
Similar to the TGSS, we find an improved fit of the dipole modulation for increasing flux density thresholds.
The strongest improvement, based on the $\chi^2/dof$, we observe between the 25 and 35~mJy flux density thresholds for both masks.

For the SUMSS $18$~mJy sample with both masks we also observe the largest amplitude across the flux density thresholds.
The measured amplitudes for the higher flux densities are in good consistency, but decrease with increasing flux density threshold.
The declination of all SUMSS samples is consistent with the CMB dipole. 
The right ascension however shows a clear offset and is within three sigma not consistent with the right ascension of the CMB dipole direction. 
We ensured, that our measurements in the SUMSS samples are not biased towards a overdense cell in the south east, which is also not masked by `mask n'.
We manually masked this cell and measured a dipole of $2.9\times10^{-3}$ towards $( 104.06, +3.58)$~deg for $S>25$~mJy and `mask d', or $3.1\times10^{-3}$ towards $( 102.66, 0.00)$~deg for $S>25$~mJy and `mask n', which is consistent with previous results.
Beside previous results, we find the largest $\chi^2/dof$ for the second flux density threshold of both masks.
The faintest samples, contrariwise to the TGSS and WENSS, show with higher flux density thresholds consistent test statistics.

The dipole amplitude results from the NVSS samples shows nearly consistently the smallest values and are self-consistent for both masks.
Beside the amplitudes reported in \citet{BlakeWall2002}, we find good agreement to the measured dipole amplitudes of \citet{Singal2011, RubartSchwarz2013, Bengaly2018} and \cite{TiwariNusser2016}.
For a detailed list on results from the NVSS see Table \ref{tab:resultsnvss}.
The measured dipole direction are in good agreement to earlier results, except for the measured direction from \citet{GibelyouHuterer2012}.
Between the two masks, we find the smallest offsets to the CMB dipole direction for `mask n' and in detail for a flux density threshold of 45~mJy.
The directional offsets of `mask d' are in good agreement to the offsets observed for the WENSS with 25 and 35~mJy flux density threshold.
The agreement of the dipole modulation to the observed source counts is for flux density thresholds larger 15~mJy consistent.
For the 15~mJy flux density threshold of both masks we find within the NVSS samples the poorest fit, which is in good agreement to previously observed systematical source density fluctuations below 15~mJy \citep{BlakeWall2002}.

In general the measurements from samples below, or close to completeness, like TGSS $50$~mJy, WENSS $25$~mJy and SUMSS $18$~mJy show inconsistent results when compared to samples with flux density thresholds above completeness.
Therefore we focus on samples with higher flux densities in the conclusion.

\subsection{TGSS and NVSS cross matched catalogue}\label{sec:crossmatch}

\begin{figure}
    \centering
    \includegraphics[width = \linewidth]{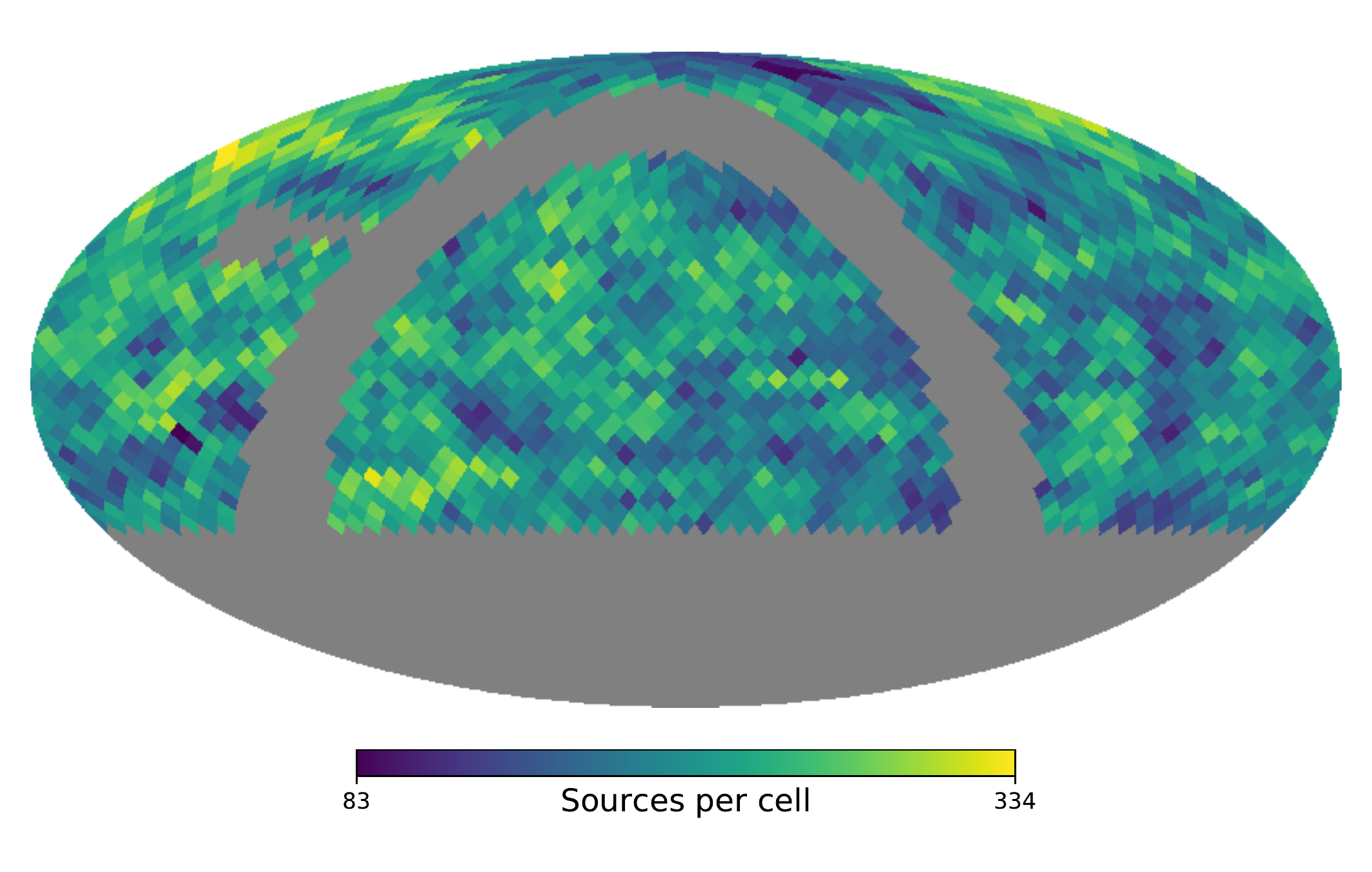}\\
    \includegraphics[width = \linewidth]{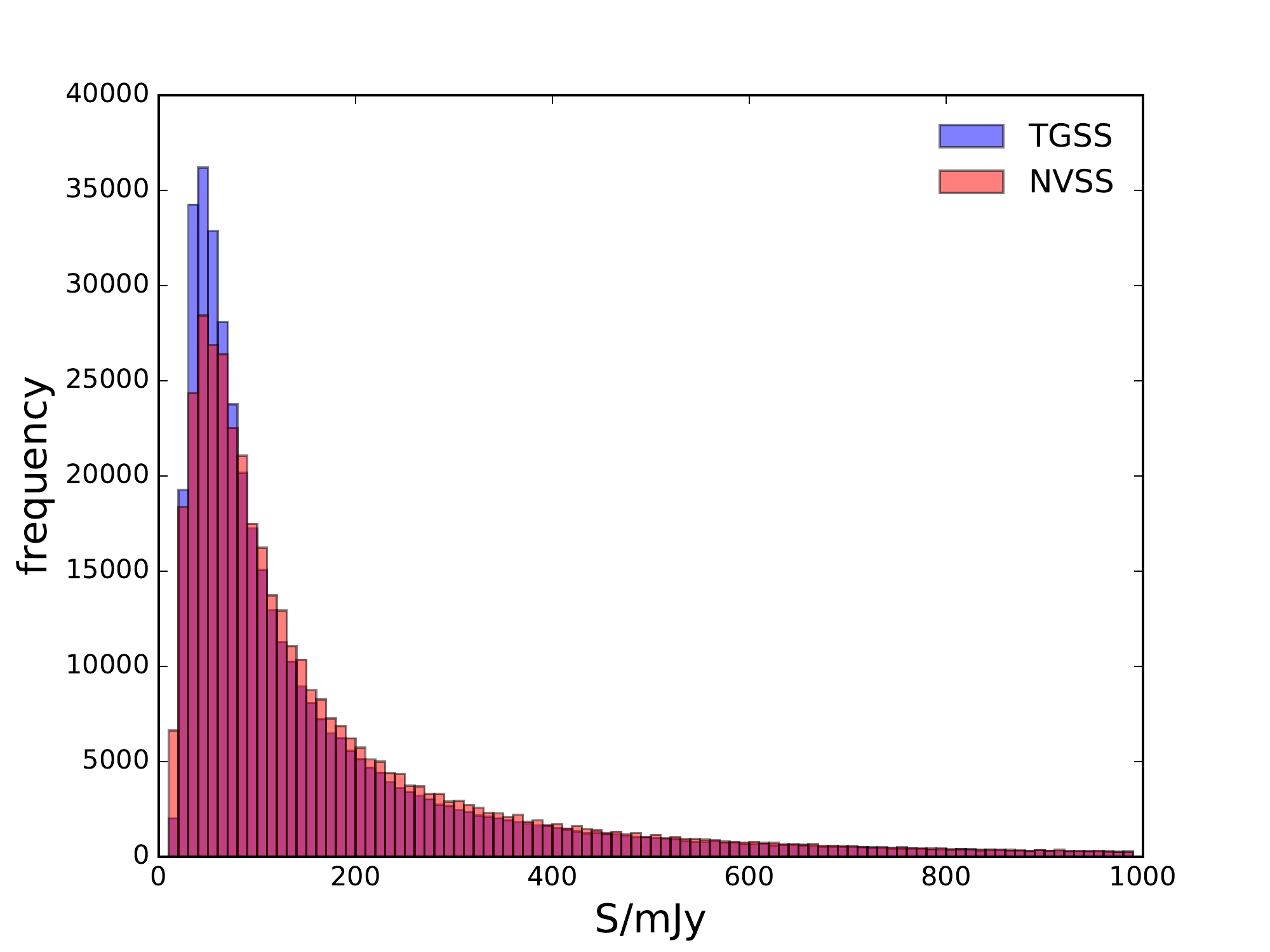}
    \caption{Masked source counts per cell of the cross matched TGSSxNVSS catalogue (top) and flux density distribution of the cross matched sources of the TGSS and NVSS survey (bottom). The flux densisites of the NVSS survey have been scaled by $\alpha = 0.75$.}
    \label{fig:crossmatched}
\end{figure}

In order to investigate the strong difference in the estimated dipole amplitude between the TGSS-ADR1 and NVSS survey, we cross match both catalogues and name it TGSSxNVSS.
The cross matching was done with {\sc topcat} \citep{TOPCAT}, where the NVSS sources are matched within $10''$ to the TGSS sources.
We generated a catalogue for all $519\,414$ cross-matched sources consisting out of TGSS and NVSS source positions, as well as flux densities at $147.5$~MHz and $1.4$~GHz. 
To mask the cross-matched catalogue properly, we combine both `mask d' of the NVSS and TGSS survey.
The masked source counts per cell of the TGSSxNVSS catalogue can be seen in Fig. \ref{fig:crossmatched} (top), with $417\,840 $ sources after masking.
We additionally show the distribution of both flux densities in Fig. \ref{fig:crossmatched} (bottom), where the NVSS flux densities have been scaled to TGSS frequency by $\alpha = 0.75$.
By eye the {\sc HEALPix} map of the TGSSxNVSS and TGSS (Fig. \ref{fig:Data}) catalogue show a similar source counts distribution across the sky and are not comparable to the homogenous source distribution of the NVSS. 
The flux density distribution of the TGSS flux densities of the TGSSxNVSS show a stronger peak and narrower distribution than the scaled flux densities of the NVSS.

\begin{table*}
\centering
\caption{Results of the Cosmic Radio Dipole for the cross-matched TGSSxNVSS catalogue. Errors are estimated from 100 bootstraps for each sample.}
\begin{tabular}{lccccccc}
\hline\hline
Survey & $S$ &$N$ & $RA$ & $DEC$ &$\Delta\theta$&  $d$& $\chi^2/dof$\\
& (mJy) & & (deg) & (deg) & (deg) & $(\times 10^{-2})$&\\\hline
TGSSxNVSS & - & $417\,840$ & $102.43\pm 3.30$ & $30.12\pm 3.84$&$72.83\pm 3.08$&$9.95\pm 0.10$&6.84\\\hline
TGSSx & 135 & $159\,240$ & $139.26\pm 4.86$ & $8.66\pm 5.70$ & $32.57\pm 4.91$ & $6.41\pm 0.59$& 1.64\\\hline
xNVSS & 25 & $176\,433$ & $131.94\pm 10.67$ & $-3.04\pm 12.56$ &$36.06\pm 10.24$& $2.05\pm 0.38$& 1.09\\\hline
\multirow{2}{*}{TGSSxNVSS} & 135 & \multirow{2}{*}{$139\,161$}& \multirow{2}{*}{$136.54\pm 5.88$} &\multirow{2}{*}{ $4.82\pm 7.62$ } & \multirow{2}{*}{$33.48\pm 5.92$}&\multirow{2}{*}{ $4.41\pm 0.54$}& \multirow{2}{*}{1.23}\\
& \& 25 & & & & & &\\\hline 
\end{tabular}
\label{tab:crossresults}
\end{table*}

We estimate the Cosmic Radio Dipole of the masked TGSSxNVSS catalogue without flux density threshold and show the result in Table \ref{tab:crossresults}.
The resulting dipole amplitude for the full TGSSxNVSS sample is the upper boundary of the assumed amplitude range in the $\chi^2$ minimisation.
Based on the large $\chi^2/dof$ in comparison to results from the former, we assume that a dipole modulation is not preferred by this sample.
As the cross-matching does not guaranty the completeness of the sample, we apply a flux density threshold to either the flux densities of the TGSS, or of the NVSS sources.
To have a similar flux density threshold in both cases, we scale the $25$~mJy flux density threshold of the NVSS sample using $\alpha = 0.75$ to the frequency of the TGSS, which is $135$~mJy.
Applying these flux density thresholds to either the TGSS, or the NVSS flux densities of the cross-matched sources we find two dipole amplitudes, which strongly differ from each other. 
The higher dipole amplitude and also dipole direction, measured for the TGSSx flux density threshold is comparable to the dipole measured for the full TGSS sample of Sect. \ref{sec:subresults}.
The same holds true for the xNVSS dipole measurement when compared to the full NVSS sample.
This suggest that the sources of both samples follow a different flux density distribution.
The $\chi^2/dof$ values of both flux density threshold samples shows a clearly improved fit of the dipole to the data than for the full TGSSxNVSS sample.
Both values are in good agreement to the corresponding samples of the original surveys, with a slight indication for a better fit in both cases.
Out of the $159\,240$ sources of the TGSSx flux density threshold sample $139\,161$ sources are the same in the xNVSS flux density threshold sample, which has initially $176\,433$ sources.
We again estimate the dipole of this common flux density threshold sample and find a dipole amplitude in between the two amplitudes of the previous results. 
The resulting dipole direction is in agreement with previous findings.

\section{Discussion}\label{sec:Discussion}
We investigated possible biases of a simple linear estimator \citep{Crawford2009}, which was used in the literature to estimate the Cosmic Radio Dipole \citep[e.g.][]{Singal2011, Singal2019, RubartSchwarz2013}.
The detailed analysis shows the limited use of such an estimator, due to directional biases introduced by the masking procedure.
Possible biases affecting the estimated dipole amplitude have already been discussed in \citet{RubartSchwarz2013}.

Therefore we analysed a quadratic estimator, based on a $\chi^2$ minimisation, which uses source counts in equal area cells to determine the Cosmic Radio Dipole. 
The quadratic estimator is tested against possible contributions from masking and resolution dependencies.
We conclude that the quadratic estimator, which was already briefly discussed in a Cosmic Radio Dipole forecast of the SKA \citep{Bengaly2019,SKARedBook2018}, is suitable to distinguish a simulated kinematic dipole of percent level from contributions of a purely random sky.
The recovered position of a simulated kinematic dipole can securely be estimated with positional offsets of 1.4~deg or 0.7~deg, for full skies with $10^6$ or $10^7$ simulated sources.
These offsets are already below the mean separation of the cells used for the pixelization of the sky, which is $3.66$~deg. 
Increasing the resolution of the search-grid for the dipole direction does not significantly improve the estimation, as it is limited by the cell size of the data.
The amplitude can be recovered with uncertainties of $\sim 10\%$ for $10^7$ simulated sources. 
A kinematic dipole with a higher amplitude is detectable for even smaller source densities.

We discussed two possible masking strategies and for cuts of equal galactic latitude we found no significant effect on the recovered dipole direction and amplitude.
Masking randomly regions of the simulated sky with $10^6$ sources can increase the directional offset from $\sim 1.4$~deg to $\sim5.5$~deg. 
The recovered dipole amplitude seems to be less affected.

We applied the quadratic estimator to the TGSS-ADR1, WENSS, SUMSS and NVSS radio source catalogues.
In general we find self-consistency for the estimated dipole amplitude and direction between two different masking strategies in each survey.
Rejecting regions with higher noise flux densities or larger positional uncertainties improves the quality of the fit, but does not significantly change the resulting Cosmic Radio Dipole.
Comparing our measured dipole directions and amplitudes to other published results of the TGSS \citep{Bengaly2018}, WENSS \citep{RubartSchwarz2013} and NVSS \citep[e.g.][]{BlakeWall2002,Singal2011, RubartSchwarz2013} shows good agreement.
The dipole amplitude, inferred from the SUMSS survey is consistent with the results from WENSS and NVSS, but shows a larger directional offset.
For a combined catalogue of NVSS and SUMSS sources (NVSUMSS), \citet{Colin2017} found a dipole with velocity of $1729\pm 187$~km s$^{-1}$ towards $(RA,DEC)=(149\pm 2, -17\pm 12)$~deg, which is in agreement to results of the NVSS.
We find the best agreement of a dipole modulation to the observed source counts, based on the test statistic $\chi^2/dof$, for the NVSS survey.
The excess of the measured dipole amplitude from WENSS and NVSS compared to the CMB still persists and increases even more for the TGSS. 
Comparing the results of NVSS dipole to the (kinematic) dipole of the CMB, we find an increased amplitude by a factor of three to four.
Contrariwise, for the more aggressive `mask n', the estimated dipole direction is in good agreement to the dipole direction of the CMB.
For the TGSS, the dipole amplitude increases by an additional factor of three to four, while compared to the NVSS.

\begin{figure}
    \centering
    \includegraphics[width=\linewidth]{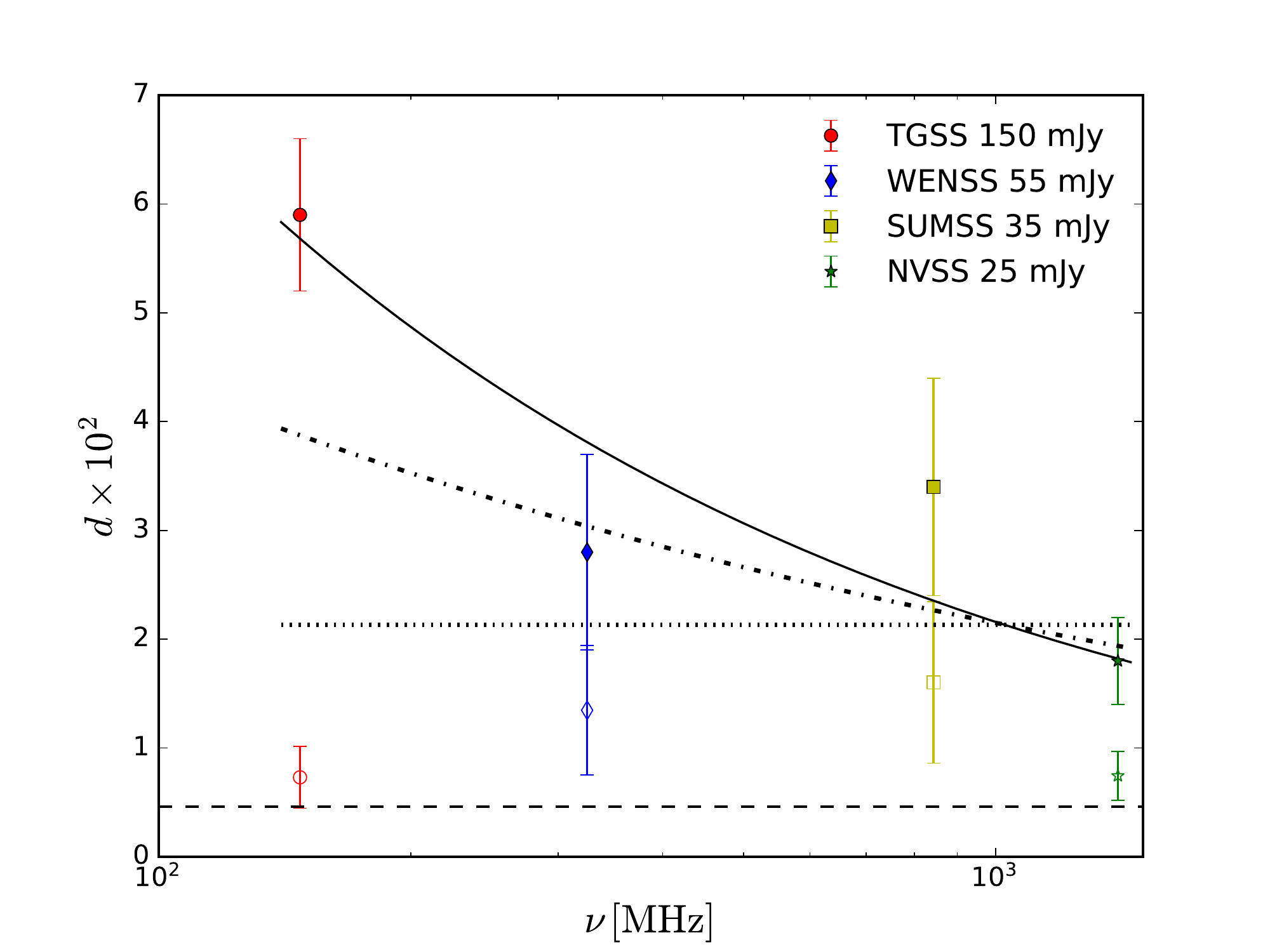}
    \caption{Comparison of dipole amplitudes observed at different frequencies for the TGSS-ADR1, WENSS, SUMSS, and NVSS masked with `mask d'. A function $f(\nu)= A(\nu/1\mathrm{~GHz})^m$ is fitted to the dipole amplitudes (solid lines), which results in $A=(2.16\pm 0.53)\times10^{-2}$ and $m=-0.51\pm 0.15$ with $\chi^2/dof= 1.23$. Results of a power law and constant dipole amplitude fit excluding the TGSS are shown as dashed-dotted and dotted lines, respectively. Simulations for each survey coverage with the CMB dipole (dashed line) are shown as open symbols.}
    \label{fig:amplitudefrequency}
\end{figure}

In order to compare the observed dipole amplitudes of each survey to the CMB dipole, we simulate source counts maps, which include the CMB dipole, the effect of limited number of sources, and reduced sky coverage.
The recovered simulated dipole amplitudes are shown in Figure \ref{fig:amplitudefrequency} with open symbols. 
We observe for the TGSS-ADR1 and NVSS a clear discrepancy between the simulated CMB dipole amplitude and the observed dipole amplitude.
Therefore the above mentioned contributions are not sufficient to explain the increased dipole amplitude with respect to the CMB dipole.
On the other hand, for the WENSS and SUMSS source catalogues, the affects of low number of sources and a decreased sky coverage could explain at least a sizable fraction of the observed excess.

\begin{figure}
    \centering
    \includegraphics[width=\linewidth]{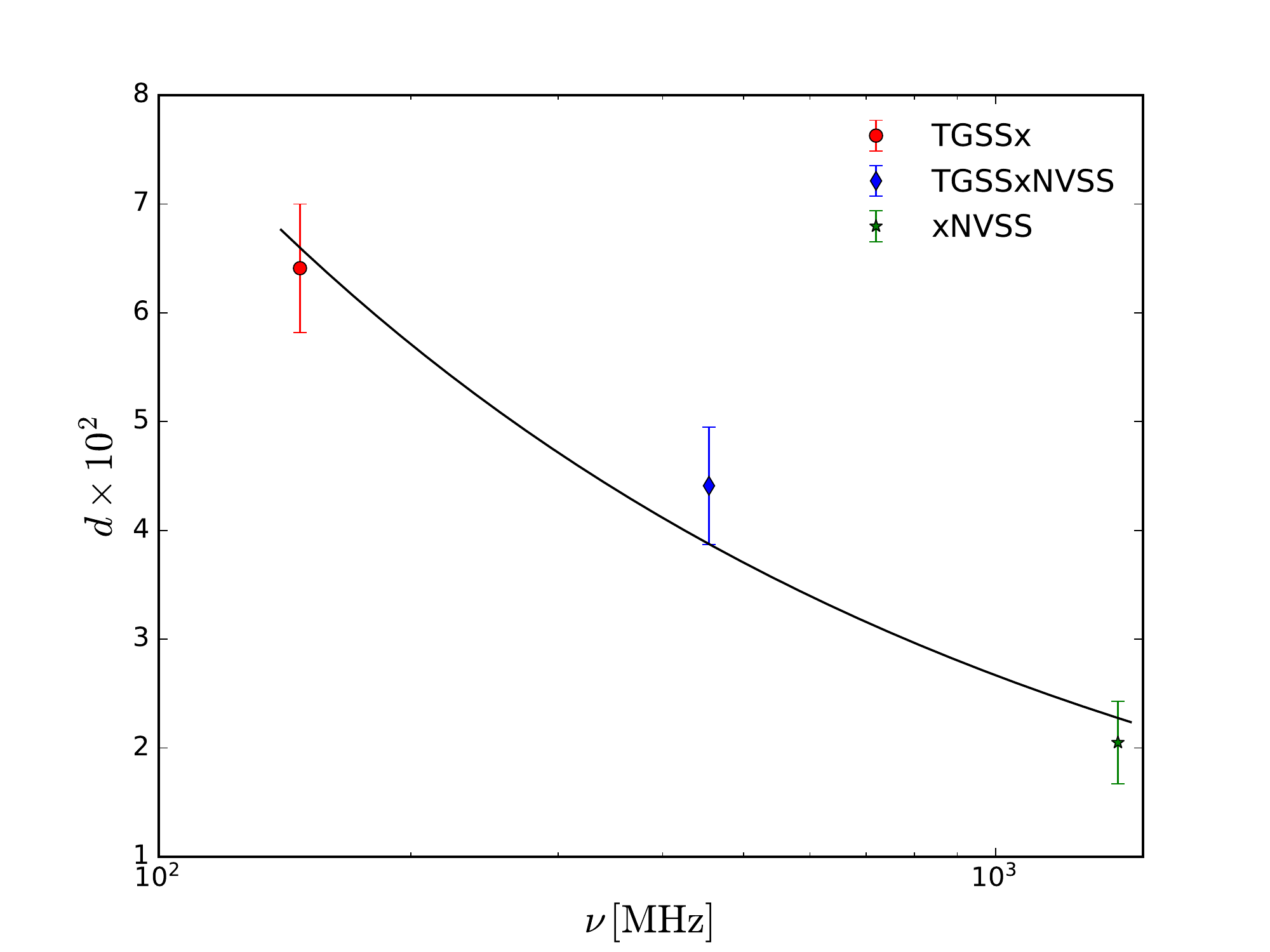}
    \caption{Comparison of dipole amplitudes from the cross-matched TGSSxNVSS catalogue. For the common flux density threshold sample of TGSSxNVSS, the geometric mean was used to determine the frequency. A linear function $f(\nu)= A(\nu/1\mathrm{~GHz})^m$ is fitted to the dipole amplitudes (solid line), which results in $A= (2.67\pm 0.70)\times10^{-2}$ and $m =-0.47\pm 0.16$, with $\chi^2/dof=1.44$.}
    \label{fig:amplitudefrequency_cross}
\end{figure}

In order to investigate the difference between the TGSS and NVSS further, we measured the dipole for a common sub-sample.
Also in the cross-matched catalogue the TGSS-ADR1 dipole amplitude is significantly larger than the NVSS dipole amplitude.
We also created a common sub-sample of the TGSSxNVSS catalogue, by scaling the NVSS flux density threshold to the TGSS frequency.
For this sub-sample, we find a intermediate dipole amplitude. 
Already for the single flux density thresholds, applied either to the TGSS or NVSS flux densities, improves the agreement of the dipole modulation to the observed source counts. 
Based on the cross-matched catalogue of TGSS and NVSS sources, we can conclude, that the observed dipole amplitudes of these two surveys are real.

The difference in the dipole amplitudes of the analysed surveys 
can not be explained by different slopes of the number counts, 
as shown in Sect.\ \ref{sec:expected}. Therefore we suggest 
a frequency dependence of the Cosmic Radio Dipole, which is 
also present in the cross-matched catalogue. 
In order to quantify the dependence of the dipole amplitude 
on the frequency, we fitted a function of the form 
$f(\nu)= A(\nu/1\mathrm{~GHz})^m$ to the dipole amplitudes 
(see Fig. \ref{fig:amplitudefrequency}).
For fitting we used the python package {\sc lmfit} 
\citep{lmfit2016} with the default Levenberg-Marquardt 
least-square minimisation. We chose the dipole amplitudes 
from our set of analysed flux density thresholds, 
such that they are roughly comparable to a flux 
density threshold scaled from 25~mJy at 1.4~GHz to the 
corresponding frequency. 
For the dipole amplitudes of `mask d', we find $A=(2.16\pm 0.53)\times10^{-2}$ and $m=-0.51\pm 0.15$ with $\chi^2/dof= 1.23$ (see Fig. \ref{fig:amplitudefrequency} solid line).
Masking with `mask n' results in a slightly steepened dependency, with $A=(1.98\pm 0.55)\times 10^{-2}$, $m=-0.61\pm 0.16$ and $\chi^2/dof=1.25$.

Excluding the TGSS result form the fit, we still find a power law in agreement with the previous fit (Fig. \ref{fig:amplitudefrequency} dashed-dotted line), 
but fitting a constant to the WENSS, SUMSS, and NVSS dipole 
amplitudes alone provides an equally good fit (Fig. \ref{fig:amplitudefrequency} dotted line). Therefore, we 
cannot exclude that the observed frequency dependence might be 
linked to issues with the flux density calibration of the TGSS 
(see the discussion below), but we see a self-consistent trend
across all frequencies.

This is also supported by a fit to dipole results of the cross-matched catalogue (see Fig. \ref{fig:amplitudefrequency_cross}). We find 
$A= (2.67\pm 0.70)\times10^{-2}$ together with $m =-0.47\pm 0.16$.
In this analysis, the dipole result of the common flux 
density threshold sample of the TGSSxNVSS catalogue was 
assigned to the geometric mean frequency of the TGSS and 
NVSS frequencies. As shown above (Table 8), a 
radio dipole dominated by the kinematic effect is 
not expected to show such a prominent dependence on frequency.

Several aspects on the dipole amplitude and direction have already been investigated in previous studies.
\citet{Rubart2014} investigated the contribution of spherically symmetric local structures to the measured dipole of a linear estimator.
By simulating a single void, while the observer is sitting at the edge of the void with radius $R_v=0.07R_H$ ($R_H$ Hubble distance), they found a dipole contribution of $d_{void}=(0.21\pm 0.01)\times10^{-2}$ for the linear estimator of \citet{RubartSchwarz2013}.
Varying the void parameters can increase the dipole contribution, but with one void it can not fully explain the discrepancy between the CMB and Radio Dipole.
Increasing the number of voids, or in general over- and under-densities would add up the contribution to the observed dipole amplitude. 
Additionally it was shown that such an contribution would be depending on the frequency of the surveys.
In general the contribution from structures in the line of sight can relax the difference between the CMB and the Radio Dipole, e.g. observed for the NVSS. 
But in order to fully explain the large discrepancy between the NVSS and TGSS dipole, one would need unrealistic huge structures.
Contributions from local structure and superclusters to the Cosmic Radio Dipole have been investigated by \citet{Colin2017} and seem to have insignificant influence on the measured large-scale dipole found in NVSS and NVSUMSS.

In a forecast of possible measurements with the upcoming SKA, \citet{Bengaly2019} simulated number count maps for the SKA survey specifications in Phase 1 and 2. 
They found a increasing contribution from local structures with increasing flux density threshold. 
Removing sources with $z\leq0.1$ from the simulations in \citet{Bengaly2019} improved the results most, which is in good agreement with results from \citet{TiwariNusser2016}.
Additional contributions, like Bulge flows, will also contribute to the Cosmic Radio Dipole.

Systematical effects, like a large-scale flux density offset for the TGSS-ADR1 compared to the GaLactic and Extragalactic All-sky Murchison Widefield Array (GLEAM; \citealt{GLEAM2017}) have been found by \citet{Tiwari2019}.
The flux density offsets are likely to contribute to the dipole and other large-scale source density measurements.
While measuring the power spectrum of the TGSS-ADR1, \citet{Tiwari2019} found a dipole of $|\vec{D}|=0.05$ at $S>100$~mJy, which is slightly smaller than previous results from \citet{Bengaly2018} and our own results.

Further distinctions between contributions from large structures and/or artefacts in the survey are necessary to reveal the real nature of the Cosmic Radio Dipole. 
To better distinguish contributions from local structure and shot noise effects with the quadratic estimator, a higher source density is needed.
Upcoming surveys, like further data releases of the LOFAR two-metre sky survey (LoTSS; \citealt{LoTSS2017}), which is carried out with the LOw Frequency ARray (LOFAR; \citealt{LOFAR2013}), will eventually cover a large enough sky fraction to measure the Cosmic Radio Dipole with much higher precision.
With follow-up studies, like the WEAVE-LOFAR survey \citep{LOFARWEAVE2016}, using the William Herschel Telescope Enhanced Area Velocity Explorer (WEAVE; \citealt{WEAVE2012, WEAVE2014}), it will be possible to measure spectroscopic redshifts for a subsample of the LoTSS survey.
As pointed out previously, forecast in terms of the SKA showed the possibility to measure the Cosmic Radio Dipole with $\sim10\%$ accuracy \citep{SKARedBook2018,Bengaly2019}.
Additionally, the Evolutionary Map of the Universe (EMU; \citealt{EMU2011}) will eventually map the entire southern sky at a frequency of 1.3~GHz and is expected to detect $\sim70$~million radio sources. 
The field of view of EMU will fully cover the Dark Energy Survey (DES; \citealt{DES2005}), which will provide photometric redshifts of optical counterparts. 
Cross-matching new radio surveys with existing optical and infrared surveys will help to distinguish contributions from local structure to the observed Cosmic Radio Dipole. 
In upcoming surveys, we expect contributions from local structure to show a similar frequency dependence, as shown in this study for the analysed surveys.

\section{Conclusions \label{sec:conclusions}}

To conclude, we analysed a quadratic estimator for the Cosmic Radio Dipole, which seems to be unbiased with respect to masking.
With this quadratic estimator we can recover estimates of the Cosmic Radio Dipole for the NVSS, WENSS and TGSS from the literature and for the first time we produced a dipole estimation for the SUMSS catalogue.

For all analysed surveys we observe an excess in the dipole amplitude with respect to the expectation, derived from the CMB dipole. The dipole direction is consistent with the direction of the CMB dipole. This results are in agreement with various other studies (e.g.\ \citealt{BlakeWall2002,Singal2011, RubartSchwarz2013}). Different estimators and masking strategies do not change this result.
We also find a frequency dependence of the observed Cosmic Radio Dipole. Such a behaviour is incompatible with a kinematic effect, which should be (almost) achromatic. However, dropping the lowest frequency survey from the analysis would allow for a frequency independent dipole amplitude, but its amplitude would acually be in conflict with the observed CMB dipole.

Upcoming telescopes and surveys, like the Square Kilometre Array, the Evolutionary Map of the Universe, and the LOFAR two-metre sky survey with higher sensitivities will hopefully help to better understand the discrepancy between the CMB and Cosmic Radio Dipole. 
In a next step a radio continuum survey complemented with photometric redshifts, like LoTSS, will allow us to test if nearby cosmic structures could be held responsible for the observed signal and its frequency dependence.

\begin{acknowledgements}

This research made use of Astropy, a community-developed core Python package for astronomy \citep{Astropy2013} hosted at http://www.astropy.org/, and of the astropy-based reproject package (\url{http://reproject.readthedocs.io/en/stable/}).
Some of the results in this paper have been derived using the healpy \citep{healpy2019} and HEALPix \citep{Healpix2005} package.
This research made use of TOPCAT \citep{TOPCAT}, matplotlib \citep{matplotlib}, NumPy \citep{NumPy}, lmfit \citep{lmfit2016} and SciPy \citep{SciPy}.
We thank the staff of the GMRT that made these observations possible. 
GMRT is run by the National Centre for Radio Astrophysics of the Tata Institute of Fundamental Research.
TMS and DJS acknowledge the Research Training Group 1620 `Models of Gravity', 
supported by Deutsche Forschungsgemeinschaft (DFG) and by the German Federal Ministry for Science and Research
BMBF-Verbundforschungsprojekt D-LOFAR IV (grant number 05A17PBA).
MSR acknowledges financial support from the Friedrich-Ebert-Stiftung.
\end{acknowledgements}

\bibliographystyle{aa} 
\bibliography{aa}

\begin{appendix}

\section{Earlier Results} \label{sec:otherresults}
In Table \ref{tab:resultsnvss}, \ref{tab:resultswenss} and \ref{tab:resultstgss} we compare the results of the Cosmic Radio Dipole for the NVSS, WENSS and TGSS-ADR1 radio source catalogues. The results have been derived from different approaches and estimators, which are shortly presented in the following.

\begin{table*}
\centering
\caption{Measurements of the Cosmic Radio Dipole in terms of the NVSS from the literature. }
\begin{tabular}{ccccccccc}
\hline \hline
Survey & $\nu$ & Reference & Estimator& $S$ & $N$& $RA$ & $DEC$ & $d$ \\
& (MHz) &&&  (mJy) & & (deg) & (deg) & $(\times10^{-2})$\\\hline
NVSS & 1400 & BW &$1\leq l\leq3$& $S>10$&$431\,990$ & $132\pm 29$ & $65\pm 19$ & $0.5\pm 0.2$ \\
&& BW &$1\leq l\leq3$& $S>15$&$311\,037$ & $148\pm 29$ & $31\pm 31$ & $0.8\pm 0.3$ \\
&& BW &$1\leq l\leq3$& $S>25$&$242\,710$ & $153\pm 27$ & $-3\pm 29$ & $1.1\pm 0.3$ \\
&& BW &$1\leq l\leq3$& $S>25$&$197\,998$ & $158\pm 30$ & $-4\pm 34$ & $1.1\pm 0.3$ \\
& & BW &$1\leq l\leq3$& $S>30$&$166\,694$ & $156\pm 32$ & $2\pm 33$ & $1.1\pm 0.4$ \\
&& BW &$1\leq l\leq3$& $S>35$&$143\,524$ & $161\pm 44$ & $-27\pm 39$ & $0.9\pm 0.4$ \\
&& BW &$1\leq l\leq3$& $S>40$&$125\,603$ & $149\pm 49$ & $-45\pm 38$ & $0.7\pm 0.5$ \\
&& S11 & source counts&  $S\geq15$ & $298\,048$ & $149\pm 9$ & $ 15\pm 9$ & $1.3\pm 0.2$\\
&& S11 & source counts&   $S\geq20$  & $229\,365$ & $158\pm 10$ & $ 2\pm 10$ & $1.4\pm 0.3$\\
&& S11 & source counts&   $S\geq25$  & $185\,474$ & $158\pm 11$ & $ -2\pm 10$ & $1.4\pm 0.3$\\
&& S11 & source counts&   $S\geq30$  & $154\,996$ & $156\pm 11$ & $ -2\pm 10$ & $1.6\pm 0.3$\\
&& S11 & source counts&   $S\geq35$  & $132\,930$ & $157\pm 11$ & $ -12\pm 11$ & $1.5\pm 0.3$\\
&& S11 & source counts&   $S\geq40$  & $115\,837$ & $158\pm 12$ & $ -19\pm 12$ & $1.4\pm 0.4$\\
&& S11 & source counts&   $S\geq50$  & $91\,957$ & $171\pm 13$ & $ -18\pm 14$ & $1.7\pm 0.4$\\
&& S11 & sky brightness& $15\leq S<1000$ & $296\,811$ & $157\pm 9$ & $-3\pm 8$ & $2.0\pm 0.5$ \\
&& S11 & sky brightness& $20\leq S<1000$ & $228\,128$ & $158\pm 10$ & $-6\pm 9$ & $2.1\pm 0.5$ \\
&& S11 & sky brightness& $25\leq S<1000$ & $184\,237$ & $159\pm 10$ & $-7\pm 9$ & $2.2\pm 0.6$ \\
&& S11 & sky brightness& $30\leq S<1000$ & $153\,759$ & $159\pm 11$ & $-7\pm 10$ & $2.2\pm 0.6$ \\
&& S11 & sky brightness& $35\leq S<1000$ & $131\,691$ & $159\pm 11$ & $-10\pm 10$ & $2.2\pm 0.6$ \\
&& S11 & sky brightness& $40\leq S<1000$ & $114\,600$ & $159\pm 12$ & $-11\pm 11$ & $2.2\pm 0.6$ \\
&& S11 & sky brightness& $50\leq S<1000$ & $90\,360$ & $163\pm 12$ & $-11\pm 11$ & $2.3\pm 0.7$ \\
&& RS & linear 3D& $S>15$ &$298\,289$ & $149\pm 18$& $-17\pm 18$ & $1.6\pm 0.5$\\
&& RS & linear 3D& $S>20$ &$229\,557$ & $153\pm 18$& $2\pm 18$ & $1.8\pm 0.6$\\
&& RS & linear 3D& $S>25$ &$185\,649$ & $158\pm 19$& $-2\pm 19$ & $1.8\pm 0.6$\\
&& RS & linear 3D& $S>30$ &$155\,120$ & $156\pm 19$& $-2\pm 19$ & $1.9\pm 0.7$\\
&& RS & linear 3D& $S>35$ &$133\,022$ & $156\pm 22$& $-11\pm 22$ & $1.9\pm 0.7$\\
&& RS & linear 3D& $S>40$ &$115\,917$ & $158\pm 26$& $-18\pm 26$ & $1.7\pm 0.8$\\
&& RS & linear 3D& $S>50$ &$91\,662$ & $170\pm 23$& $-17\pm 23$ & $2.0\pm 0.8$\\
&& RS & linear 2DCG & $S>10$ & $447\, 459$ & $133\pm 14$ & -- & $1.5\pm 0.4$\\
&& RS & linear 2DCG & $S>15$ & $313\, 724$ & $148\pm 14$ & -- & $1.5\pm 0.4$\\
&& RS & linear 2DCG & $S>20$ & $241, 399$ & $150\pm 14$ & -- & $1.8\pm 0.5$\\
&& RS & linear 2DCG & $S>25$ & $192\,245$ & $155\pm 14$ & -- & $1.9\pm 0.5$\\
&& RS & linear 2DCG & $S>30$ & $163\, 208$ & $153\pm 15$ & -- & $2.0\pm 0.6$\\
&& RS & linear 2DCG & $S>35$ & $139\, 851$ & $152\pm 17$ & -- & $1.9\pm 0.6$\\
&& RS & linear 2DCG & $S>40$ & $121\, 831$ & $146\pm 20$ & -- & $1.8\pm 0.7$\\
&& RS & linear 2DCG & $S>50$ & $96\,337$ & $171\pm 19$ & -- & $1.9\pm 0.7$\\
&& B18 & hemispheres & $20<S<1000$  & $253\,313$ & $148\pm 11$&$-17\pm3$&$2.3\pm0.4$\\
&& GH & & 15& $211\,487$ & $117\pm 20$ & $6\pm 14$ & $2.7\pm0.5$\\
&& TN16 & $C_1$ & $15<S<1000$ & & $151$ & $-6$ & $0.9\pm 0.4$\\
&& TN16 & $C_1$& $20<S<1000$ & & $151$ & $-14$ & $1.2\pm 0.5$\\
&& TN16 & $C_1$& $30<S<1000$ & & $160$ & $-16$ & $1.3\pm 0.6$\\
&& TN16 & $C_1$& $40<S<1000$ & & $150$ & $-35$ & $1.4\pm 0.7$\\
&& TN16 & $C_1$& $50<S<1000$ & & $174$ & $-36$ & $1.7\pm 0.8$\\\hline
\end{tabular}	
\label{tab:resultsnvss}
\tablebib{ BW:~\citet{BlakeWall2002}; S11:~\citet{Singal2011}; RS:~\citet{RubartSchwarz2013}; B18:~\citet{Bengaly2018}; GH:~\citet{GibelyouHuterer2012}; TN16:~\citet{TiwariNusser2016}}
\end{table*}
\begin{table*}
\centering
\caption{Measurements of the Cosmic Radio Dipole in terms of the WENSS from the literature.}
\begin{tabular}{ccccccccc}
\hline \hline
Survey & $\nu$ & Reference & Estimator& $S$ & $N$& $RA$ & $DEC$ & $d$ \\
& (MHz) &&&  (mJy) & & (deg) & (deg) & $(\times10^{-2})$\\\hline
WENSS & 325& RS & linear 2DCG & 25& $92\,600$ & $117\pm 40$& -- & $2.9\pm 1.9$\\
&& RS & linear 2DCG & 30& $92\,600$ & $122\pm 44$& -- & $2.8\pm 2.1$\\
&& RS & linear 2DCG & 35& $92\,600$ & $123\pm 46$& -- & $2.9\pm 2.1$\\
&& RS & linear 2DCG & 40& $92\,600$ & $124\pm 51$& -- & $2.9\pm 2.3$\\\hline
\end{tabular}
\label{tab:resultswenss}
\tablebib{RS:~\cite{RubartSchwarz2013}}
\end{table*}

\begin{table*}
\centering
\caption{Measurements of the Cosmic Radio Dipole in terms of the TGSS-ADR1 from the literature.}
\begin{tabular}{ccccccccc}
\hline \hline
Survey & $\nu$ & Reference & Estimator& $S$ & $N$& $RA$ & $DEC$ & $d$ \\
& (MHz) &&&  (mJy) & & (deg) & (deg) & $(\times10^{-2})$\\\hline
TGSS & 147& B18 & hemispheres & $100<S<5000$&$233\,395$ & $155\pm 12$ & $0\pm 3$ & $7.0\pm 0.4$ \\
&&S19 & source counts& $S\geq 100$ &$227\,773$ & $162 \pm 9$& $3\pm 8$& $4.42\pm0.37$ \\
&&S19 & source counts& $S\geq 150$ &$161\,664$ & $164 \pm 9$& $-1\pm 8$& $4.65\pm0.42$ \\
&&S19 & source counts& $S\geq 200$ &$124\,080$ & $166 \pm 10$& $-3\pm 9$& $4.65\pm0.45$ \\
&&S19 & source counts& $S\geq 250$ &$99\,736$ & $168 \pm 10$& $-8\pm 9$& $5.02\pm0.50$ \\
&&S19 & sky brightness& $100\leq S<5000$ &$226\,242$ & $172 \pm 10$& $-5\pm 9$& $5.24\pm0.53$ \\
&&S19 & sky brightness& $150\leq S<5000$ &$160\,133$ & $173 \pm 10$& $-6\pm 9$& $5.39\pm0.57$ \\
&&S19 & sky brightness& $200\leq S<5000$ &$122\,549$ & $174 \pm 11$& $-6\pm 10$& $5.43\pm0.60$ \\
&&S19 & sky brightness& $250\leq S<5000$ &$98\,205$ & $174 \pm 11$& $-8\pm 10$& $5.64\pm0.64$ \\\hline
\end{tabular}	
\label{tab:resultstgss}
\tablebib{ B18:~\cite{Bengaly2018}; S19:~\cite{Singal2019}}
\end{table*}

\citet{RubartSchwarz2013} uses a three dimensional linear estimator (\textit{linear 3D}), introduced by \citet{Crawford2009}:
\begin{equation}\label{eq:3Dlinearestimator}
R_{3D}=\sum_i^N \hat{r}_i,
\end{equation}
which points towards the main anisotropy of the source distribution with sources at position $r_i$.
As the masking affects the measured dipole amplitude, they correct the estimated dipole amplitude $d=|\frac{3}{N}R_{3D}|$ like:
\begin{equation}
    \langle d_{3D}\rangle_{\text{mask}} = \sqrt{k_{3D}^2d^2+\frac{9}{N^2}\langle R_{3D}^{\text{random}}\rangle},
\end{equation}
with a correction factor estimated from simulations with a given dipole:
\begin{equation}
k = \sqrt{\tilde{k}^2+\frac{9\langle R^{Random}\rangle^2}{d^2N^2}(\tilde{k}^2-1)}.
\end{equation}
Due to directional bias in the declination, \citet{RubartSchwarz2013} introduced in the same work a two-dimensional estimator (\textit{linear 2DCG}):
\begin{equation}
    R_{2D} = \sum_i^N \begin{pmatrix}\cos\varphi_i\sin\vartheta_i\\\sin\varphi_i\sin\vartheta_i\\0\end{pmatrix},
\end{equation}
with a masking correction for the dipole amplitude:
\begin{equation}
\langle d_{2D} \rangle_{\text{mask}}=\sqrt{k^2_{2D}d^2\sin^2\vartheta_d+\frac{9}{N^2}\langle R^{\text{random}_{2D}}\rangle ^2}.
\end{equation}
\citet{Singal2011, Singal2019} uses the same three-dimensional estimator (Eq. \ref{eq:3Dlinearestimator}) to estimate the direction of the dipole, while the dipole amplitude is estimated via the difference of the \textit{source counts}:
\begin{equation}
d = \frac{3}{2k}\frac{\Delta N}{N}=\frac{3}{2k}\frac{\sum \cos \theta_i}{\sum |\cos\theta_i|}
\end{equation}
with $\theta_i$ the polar angle between the $i$-th source and the dipole direction, and a numerical computed correction factor $k$.
Similarly, \citet{Singal2011, Singal2019} uses the flux density to define a \textit{sky brightness} estimator for the dipole direction:
\begin{equation}
    R=\sum_i^N S_i \hat{r}_i.
\end{equation}
The dipole amplitude can then be estimated via:
\begin{equation}
d =  \frac{3}{2k}\frac{\Delta F}{F}=\frac{3}{2k}\frac{\sum S_i \cos \theta_i}{\sum S_i |\cos \theta_i|}.
\end{equation}
\citet{GibelyouHuterer2012} compares {\sc HEALPix} source count maps to maps of randomly chosen directions in order to find the dipole direction via:
\begin{equation}
R_{3DM}= \sum_{i}^{N_D} \hat{r_i}-\frac{N_D}{N_R}\sum_{j}^{N_R}\hat{r_j}
\end{equation}
The dipole amplitude is then estimated from the likelihood function marginalised over all directions.

Based on a {\sc HEALPix} source count map, \citet{TiwariNusser2016} convert the density contrast
\begin{equation}
\Delta(\hat{r})= \frac{N(\hat{r})}{\bar{N}}-1
\end{equation}
into spherical harmonics:
\begin{equation}
\Delta(\hat{r})=\sum_{l=1}^{\infty}\sum_{m=-l}^{+l}a_{lm}Y_{lm}(\hat{r}),
\end{equation}
\begin{equation}
a_{lm}=\int_{4\pi} d\Omega\Delta(\hat{r})Y_{lm}(\hat{r}).
\end{equation}
Using the observed power spectrum:
\begin{equation}
C_l^{obs}= \frac{\langle |a_{lm}|^2\rangle}{J_{lm}}- \frac{1}{\bar{N}},
\end{equation}
of the observed density contrast, with a correction for the survey coverage:
\begin{equation}
J_{lm}= \int_{survey}d\Omega|Y_{lm}|^2,
\end{equation}
they compute the dipole amplitude like:
\begin{equation}
C_1^{obs}= \frac{4\pi}{9}d^2.
\end{equation}

Due to the incomplete sky coverage, \citet{BlakeWall2002} measured moments $1\leq l\leq 3$ of the spherical harmonics.
In their work the dipole amplitude and direction are estimated by varying both on a given grid and compare them to the observed harmonic coefficients:
\begin{equation}
    a_{l,m} = \sum_{i=1}^NY^*_{l,m}(\vartheta_i,\varphi_i).
\end{equation}

Another approach to use the source counts defined by a {\sc HEALPix} map is investigated by \citet{Bengaly2018}.
They split the sky into hemispheres and calculate the source count difference of opposite hemispheres via:
\begin{equation}
    \Delta(\theta) = \frac{\sigma_i^U(\theta)-\sigma_i^D(\theta)}{\sigma}= d_{obs}\cos\theta,
\end{equation}
with $\theta$ the separation of the estimated direction to the $i$-th pixel and the source density of the `up' (U) or `down' (D) hemispheres:
\begin{equation}
    \sigma_i=\frac{N_i}{2\pi(f_{sky})_i }.
\end{equation}

\end{appendix}

\end{document}